\documentclass[journal]{IEEEtran}
\usepackage[noadjust]{cite}

\usepackage[caption=false]{subfig}
\usepackage{graphicx}
\usepackage{amsmath}
\usepackage{multirow}
\usepackage{siunitx}
\usepackage{amssymb, textcomp}
\usepackage[hyphens]{url}
\usepackage{dblfloatfix}
\usepackage{xcolor}

\begin{document}
\title{Sound Event Localization and Detection \\of Overlapping Sources Using\\ Convolutional Recurrent Neural Networks}
\author{Sharath~Adavanne,~\IEEEmembership{Member,~IEEE,}
        Archontis~Politis,~\IEEEmembership{Member,~IEEE,}
        Joonas~Nikunen,~\IEEEmembership{Member,~IEEE,}
        and~Tuomas Virtanen,~\IEEEmembership{Member,~IEEE}
\thanks{S. Adavanne, J. Nikunen and T.Virtanen are with the Signal Processing Laboratory, Tampere University of Technology, Finland, e-mail: firstname.lastname@tut.fi}
\thanks{A. Politis is with the Department of Signal Processing and Acoustics, Aalto University, Finland, e-mail: archontis.politis@aalto.fi}
\thanks{The research leading to these results has received funding from the European Research Council under the European Union’s H2020 Framework Program through ERC Grant Agreement 637422 EVERYSOUND. The authors also wish to acknowledge CSC-IT Center for Science, Finland, for computational resources}
\thanks{This work has been submitted to the IEEE for possible publication. Copyright may be transferred without notice, after which this version may no longer be accessible.}
}

%

\maketitle
\bstctlcite{IEEEexample:BSTcontrol}
\begin{abstract}
In this paper, we propose a convolutional recurrent neural network for joint sound event localization and detection (SELD) of multiple overlapping sound events in three-dimensional (3D) space. The proposed network takes a sequence of consecutive spectrogram time-frames as input and maps it to two outputs in parallel. As the first output, the sound event detection (SED) is performed as a multi-label classification task on each time-frame producing temporal activity for all the sound event classes. As the second output, localization is performed by estimating the 3D Cartesian coordinates of the direction-of-arrival (DOA) for each sound event class using multi-output regression. The proposed method is able to associate multiple DOAs with respective sound event labels and further track this association with respect to time. The proposed method uses separately the phase and magnitude component of the spectrogram calculated on each audio channel as the feature, thereby avoiding any method- and array-specific feature extraction. The method is evaluated on five Ambisonic and two circular array format datasets with different overlapping sound events in anechoic, reverberant and real-life scenarios. The proposed method is compared with two SED, three DOA estimation, and one SELD baselines. The results show that the proposed method is generic and applicable to any array structures, robust to unseen DOA values, reverberation, and low SNR scenarios. The proposed method achieved a consistently higher recall of the estimated number of DOAs across datasets in comparison to the best baseline. Additionally, this recall was observed to be significantly better than the best baseline method for a higher number of overlapping sound events.

\end{abstract}


\IEEEpeerreviewmaketitle

\section{Introduction}

\IEEEPARstart{S}{ound} event localization and detection (SELD) is the combined task of identifying the temporal activities of each sound event, estimating their respective spatial location trajectories when active, and further associating textual labels with the sound events. Such a method can for example automatically describe social and human activities and assist the hearing impaired to visualize sounds. Robots can employ this for navigation and natural interaction with surroundings~\cite{Takeda2016_single,Takeda2016_double,Yalta2017,He2018}. Smart cities, smart homes, and industries could use it for audio surveillance~\cite{surveillance_audio, Grobler2017, Wessels2017, Foggia_TITS2015}. Smart meeting rooms can recognize speech among other events and use this information to beamform and enhance the speech for teleconferencing or for robust automatic speech recognition~\cite{busso2005smart, Wang1997,Woelfel2009,ghoshal2014,Butko2011}. Naturalists could use it for bio-diversity monitoring~\cite{environmentalSED,Marques2012,Furnas2014}. Further, in virtual reality (VR) applications with \ang{360} audio SELD can be used to assist the user in visualizing sound events.

\begin{table*}[!ht]
\centering
\caption{Summary of DNN based DOA estimation methods in the literature. The azimuth and elevation angles are denoted as `azi' and `ele', distance as `dist', `x' and `y' represent the distance along the respective Cartesian axis. `Full' represents the estimation in the complete range of the respective format, and `regression' represents the classifier estimation type.}
\label{T:doa_methods_summary}
\begin{tabular}{lllcccc}
Approach & Input feature & Output format & Sources & DNN & Array & SELD \\ & \\ [-1.5ex] \hline & \\ [-1.5ex]
Chakrabarty et al.~\cite{Chakrabarty2017,Chakrabarty2017_nips} & Phase spectrum & azi & 1, multiple & CNN & Linear & $\times$ \\
Yalta et al.~\cite{Yalta2017} & Spectral power & azi (Full) & 1 & CNN Resnet & Robot & $\times$ \\
Xiao et al.~\cite{Xiao2015} & GCC & azi (Full) & 1 & FC & Circular & $\times$\\
Takeda et al.~\cite{Takeda2016_single,Takeda2016_double} & \begin{tabular}[c]{@{}l@{}}Eigen vectors of spatial \\ covariance matrix\end{tabular} & azi (Full) & 1, 2 & FC & Robot & $\times$ \\
He et al.~\cite{He2018}& GCC & azi (Full) & Multiple & CNN  & Robot & $\times$ \\
Hirvonen~\cite{Hirvonen2015} & Spectral power & azi (Full) for each class  & Multiple & CNN & Circular & \checkmark \\ 
Yiwere et al.~\cite{Yiwere2017} & ILD, cross-correlation & azi and dist & 1 & FC & Binaural & $\times$ \\
Ferguson et al.~\cite{Ferguson2017} & GCC, cepstrogram & azi and dist (regression) & 1 & CNN & Linear & $\times$ \\
Vesperini et al.~\cite{Vesperini2016} & GCC & x and y (regression) & 1 & FC & Distributed & $\times$ \\
Sun et al.~\cite{Sun2017} & GCC & \begin{tabular}[c]{@{}l@{}}azi and ele\end{tabular} & 1 & PNN & Cartesian & $\times$ \\
Adavanne et al.~\cite{Adavanne2018_EUSIPCO}& Phase and magnitude spectrum & azi and ele (Full) & Multiple & CRNN & Generic & $\times$ \\
Roden et al.~\cite{Roden2015} & \begin{tabular}[c]{@{}l@{}}ILD, ITD, phase and \\ magnitude spectrum\end{tabular} & \begin{tabular}[c]{@{}l@{}}azi, ele and dist \\ (separate NN)\end{tabular} & 1 & FC & Binaural & $\times$ \\  & \\ [-1.5ex] \hline & \\ [-1.5ex]
Proposed & Phase and magnitude spectrum & \begin{tabular}[c]{@{}l@{}} azi and ele (Full, \\regression) for each class \end{tabular} & Multiple & CRNN &  Generic & \checkmark
\end{tabular} 
\end{table*}

\subsection{Sound event detection}
The SELD task can be broadly divided into two sub-tasks, sound event detection (SED) and sound source localization. SED aims at detecting temporally the onsets and offsets of sound events and further associating textual labels to the detected events. The sound events in real-life most often overlap with other sound events in time and the task of recognizing all the overlapping sound events is referred as polyphonic SED. The SED task in literature has most often been approached using different supervised classification methods that predict the framewise activity of each sound event class. Some of the classifiers include Gaussian mixture model (GMM) - hidden Markov model (HMM)~\cite{Mesaros2010_EUSIPCO}, fully connected (FC) neural networks~\cite{emre2015}, recurrent neural networks (RNN)~\cite{giam2016,Adavanne_DCASE2016,Hayashi_TASLP2017,Zohrer_INTERSPEECH2017}, and convolutional neural networks (CNN)~\cite{Zhang2015,Phan2016}. More recently state-of-the-art results were obtained by stacking CNN, RNN and FC layers consecutively, referred jointly as the convolutional recurrent neural network (CRNN)~\cite{Adavanne2018_IJCNN,Lim2017,emre_TASLP2016,Adavanne_DCASE2017_binaural,Adavanne2017}. 

Lately, in order to improve recognition of overlapping sound events, several multichannel SED methods have been proposed~\cite{Adavanne2017,Jeong2017,Zhou2017,Lu2017,temko2007} and these were among the top performing methods in the real-life SED task of DCASE 2016\footnote{\url{http://www.cs.tut.fi/sgn/arg/dcase2016/task-results-sound-event-detection-in-real-life-audio#system-characteristics}} and 2017\footnote{\url{http://www.cs.tut.fi/sgn/arg/dcase2017/challenge/task-sound-event-detection-in-real-life-audio-results#system-characteristics}} evaluation challenges. More recently, we studied the SED performance on identical sound scenes captured using single, binaural and first-order Ambisonics (FOA) microphones~\cite{Adavanne2018_IJCNN}, where the order denotes the spatial resolution of the format and the first order corresponds to four channels. The results showed that the recognition of overlapping sound events improved with increase in spatial sampling, and the best performance was obtained with FOA. 



\subsection{Sound source localization}

Sound source localization is the task of determining the direction or position of a sound source with respect to the microphone. In this paper, we only deal with the estimation of the sound event direction, generally referred as direction-of-arrival (DOA) estimation. The DOA methods in literature can be broadly categorized into parametric- and deep neural network (DNN)-based approaches. Some popular parametric methods are based on time-difference-of-arrival (TDOA)~\cite{Huang2001}, the steered-response-power (SRP)~\cite{Brandstein1997}, multiple signal classification (MUSIC)~\cite{Schmidt1986}, and the estimation of signal parameters via rotational invariance technique (ESPRIT)~\cite{Roy1989}. These methods vary in terms of algorithmic complexity, constraints in array geometry, and model assumptions on the acoustic scenarios. Subspace methods like MUSIC can be applied with different array types and can produce high-resolution DOA estimates of multiple sources. On the other hand, subspace methods require a good estimate of the number of active sources that may be hard to obtain, and they have been found sensitive to reverberant and low signal-to-noise (SNR) scenarios~\cite{dibiase2001robust}. 

Recently, DNN-based methods were employed to overcome some of the drawbacks of parametric methods, while being robust towards reverberation and low SNR scenarios. Additionally, implementing the localization task in the DNN framework allows seamless integration into broader DNN tasks such as SELD~\cite{Hirvonen2015}, robots can use it for sound source based navigation and natural interaction in multi-speaker scenarios~\cite{Takeda2016_single,Takeda2016_double,Yalta2017,He2018}. A summary of the most recent DNN-based DOA estimation methods is presented in Table~\ref{T:doa_methods_summary}. All these methods estimate DOAs for static point sources and were shown to perform equally or better than the parametric methods in reverberant scenarios. Further, methods~\cite{Chakrabarty2017_nips,He2018,Adavanne2018_EUSIPCO,Hirvonen2015} proposed to simultaneously detect DOAs of overlapping sound events by estimating the number of active sources from the data itself. Most methods used a classification approach, thereby estimating the source presence likelihood at a fixed set of angles, while~\cite{Vesperini2016,Ferguson2017} used a regression approach and let the DNN produce continuous output. 

All of the past works were evaluated on different array geometries, making a direct performance comparison difficult. Most of the methods estimated full azimuth ('Full' in Table~\ref{T:doa_methods_summary}) using microphones mounted on a robot, circular and distributed arrays, while the rest of the methods used linear arrays thereby estimating only the azimuth angles in a range of $\ang{180}$. Although few of the existing methods estimated the azimuth and elevation jointly~\cite{Sun2017,Adavanne2018_EUSIPCO}, most of them estimated only the azimuth angle~\cite{Takeda2016_double, Xiao2015, Takeda2016_single, Chakrabarty2017, Chakrabarty2017_nips, Yalta2017, He2018,Hirvonen2015}. In particular, we studied the joint estimation of azimuth and elevation angles in~\cite{Adavanne2018_EUSIPCO}, this was enabled by the use of Ambisonic signals (FOA) obtained using a spherical array. Ambisonics are also known as spherical harmonic (SH) signals in the array processing literature, and they can be obtained from various array configurations such as circular or planar (for 2D capture) and spherical or volumetric (for 3D capture) using an appropriate linear transform of the recordings~\cite{teutsch2007modal}. The same ambisonic channels have the same spatial characteristics independent of the recording setup, and hence, studies on such hardware-independent formats make the evaluation and results more easily comparable in the future. 

Most of the previously proposed DNN-based DOA estimation methods that relied on a single array or distributed arrays of omnidirectional microphones, captured source location information mostly in phase- or time-delay differences between the microphones. However, compact microphone arrays with full azimuth and elevation coverage, such as spherical microphone arrays, rely strongly on the directionality of the sensors to capture spatial information, this reflects mainly in the magnitude differences between channels. Motivated by this fact we proposed to use both the magnitude and phase component of the spectrogram as input features in~\cite{Adavanne2018_EUSIPCO}. Thus making the DOA estimation method~\cite{Adavanne2018_EUSIPCO} generic to array configuration by avoiding method-specific feature extractions like inter-aural level difference (ILD), the inter-aural time difference (ITD), generalized cross-correlation (GCC) or eigenvectors of spatial covariance matrix used in previous methods (Table~\ref{T:doa_methods_summary}).

\subsection{Joint localization and detection}

In the presence of multiple overlapping sound events, the DOA estimation task becomes the classical tracking problem of associating correctly the multiple DOA estimates to respective sources, without necessarily identifying the source~\cite{valin2007robust, traa2014multiple}. The problem is further extended for the polyphonic SELD task if the SED and DOA estimation are done separately, resulting in the data association problem between the recognized sound events and the estimated DOAs~\cite{Butko2011}. One solution to the data association problem is to jointly predict the SED and DOA. In this regard, to the best of the authors' knowledge,~\cite{Hirvonen2015} is the only DNN-based method which performs SELD.  Other works combining SED and parametric DOA estimation include~\cite{Grobler2017,Butko2011, Chakraborty_ICASSP2014, Lopatka2016}. Lopatka et al.~\cite{Lopatka2016} used a 3D sound intensity acoustic vector sensor, with MPEG-7 spectral and temporal features along with a support vector machine classifier to estimate DOA along azimuth for five classes of non-overlapping sound events. Butko et al.~\cite{Butko2011} used distributed microphone arrays to recognize 14 different sound events with an overlap of two at a time, using a GMM-HMM classifier, and localized them inside a meeting room using the SRP method. Chakraborty et al.~\cite{Chakraborty_ICASSP2014} replaced SRP-based localization in~\cite{Butko2011} with a sound-model-based localization, thereby fixing the data association problem faced in~\cite{Butko2011}. In contrast, Hirvonen~\cite{Hirvonen2015}, extracted the frame-wise spectral power from each microphone of a circular array and used a CNN classifier to map it to eight angles in full azimuth for each sound event class in the dataset. In this output format, the resolution of azimuth is limited to the trained directions and the performance of unseen DOA values is unknown. For larger datasets with a higher number of sound events and increased resolution along azimuth and elevation directions, this approach results in a large number of output nodes. Training such a DNN with a large number of output nodes where the number of positive class labels per frame is one or two with respect to a high number of negative class labels poses challenges of an imbalanced dataset. Additionally, training such a large number of classes requires a huge dataset with enough examples for each class. On the other hand, this output format allows the network to simultaneously recognize more than one instance of the same sound event in a given time frame, at different locations.

\subsection{Contributions of this paper}
In general, the number of existing SELD methods is limited~\cite{Grobler2017,Butko2011,Hirvonen2015,Chakraborty_ICASSP2014, Lopatka2016}, with only one published DNN-based approach~\cite{Hirvonen2015}. On the other hand, there are several DNN-based methods in the literature for the SELD sub-tasks of SED and DOA estimation. Yet, there is no comprehensive work published that studies the various choices affecting the performance of these DNN-based SED, DOA and SELD methods, compare them with multiple competitive baselines, and evaluate them over a wide range of acoustic conditions. Besides, with respect to the SELD task, the existing methods~\cite{Grobler2017,Butko2011, Chakraborty_ICASSP2014, Lopatka2016} localize up to one or maximum two overlapping sound events and do not scale to a higher number of overlapping sources. Further, the only DNN-based SELD method~\cite{Hirvonen2015} localizes sound events exclusively at a predefined grid of directions and requires a large number of output classes for a higher number of sound event labels and increased spatial resolution. Additionally, all the above SELD approaches use method-specific features and hence not independent of input array structure.

In contrast to existing SELD methods, this paper presents novelty in two broad areas: the proposed SELD method, and the exhaustive evaluation studies presented. The novelty of the proposed SELD method is as follows. It is the first method that addresses the problem of localizing and recognizing more than two overlapping sound events simultaneously and tracking their activity with respect to time. The proposed method is able to localize sources at any azimuth and elevation angles while being robust to unseen spatial locations, reverberation, and ambiance. Further, the method itself is generic enough to learn to perform SELD from any input array structure. Specifically, as our method, we propose to use the polyphonic SED output~\cite{Adavanne2017} as a confidence measure for choosing the DOAs estimated in a regression manner. By this approach, we not only extend the state-of-the-art polyphonic SED performance~\cite{Adavanne2017} for polyphonic SELD but also tackle the data-association problem faced due to the polyphony in SELD tasks~\cite{Butko2011}. As the second broad area of novelty, we present the performance of the proposed method with respect to various design choices made such as the DNN architecture, input feature and DOA output format. Additionally, we also present the comprehensive results of the proposed method with respect to six baselines (two SED, three DOA estimation, and one SELD baseline) evaluated on seven datasets with different acoustic conditions (anechoic and reverberant scenarios with simulated and real-life impulse responses), array configurations (Ambisonic and circular array) and the number of overlapping sound events.

In order to facilitate reproducibility of research, the proposed method and all the datasets used have been made publicly available\footnote{https://github.com/sharathadavanne/seld-net}. Additionally, the real-life impulse responses used to simulate datasets have also been published to enable users to experiment with custom sound events.

The rest of the paper is organized as follows. In Section~\ref{sec:method}, we describe the proposed SELD method and the training procedure. In Section~\ref{sec:evaluation}, we describe the datasets, the baseline methods, the metrics and the experiments carried out for evaluating the proposed method. The experimental results on the evaluation datasets are presented, compared with baselines and discussed in Section~\ref{sec:results}. Finally, we summarize the conclusions of the work in Section~\ref{sec:conclusion}.

\section{Method} 
\label{sec:method}
The block diagram of the proposed method for SELD is presented in Figure~\ref{fig:crnn}. The input to the method is the multichannel audio. The phase and magnitude spectrograms are extracted from each audio channel and are used as separate features. The proposed method takes a sequence of features in consecutive spectrogram frames as input and predicts all the sound event classes active for each of the input frames along with their respective spatial location, producing the temporal activity and DOA trajectory for each sound event class. In particular, a CRNN is used to map the feature sequence to the two outputs in parallel. At the first output, SED is performed as a multi-label classification task, allowing the network to simultaneously estimate the presence of multiple sound events for each frame. At the second output, DOA estimates in the continuous 3D space are obtained as a multi-output regression task, where each sound event class is associated with three regressors that estimate the 3D Cartesian coordinates $x$, $y$ and $z$ of the DOA on a unit sphere around the microphone. The SED output of the network is in the continuous range of [0 1] for each sound event in the dataset, and this value is thresholded to obtain a binary decision for the respective sound event activity as shown in Figure~\ref{fig:crnn_output}. Finally, the respective DOA estimates for these active sound event classes provide their spatial locations. The detailed description of the feature extraction and the proposed method is explained in the following sections.

\subsection{Feature extraction}
\begin{figure}[htp]
  \centering  
     \subfloat[SELDnet]{{\includegraphics[height=16cm,width=\linewidth,keepaspectratio, trim=0.1cm 0.3cm 0.1cm 0.1cm,clip]{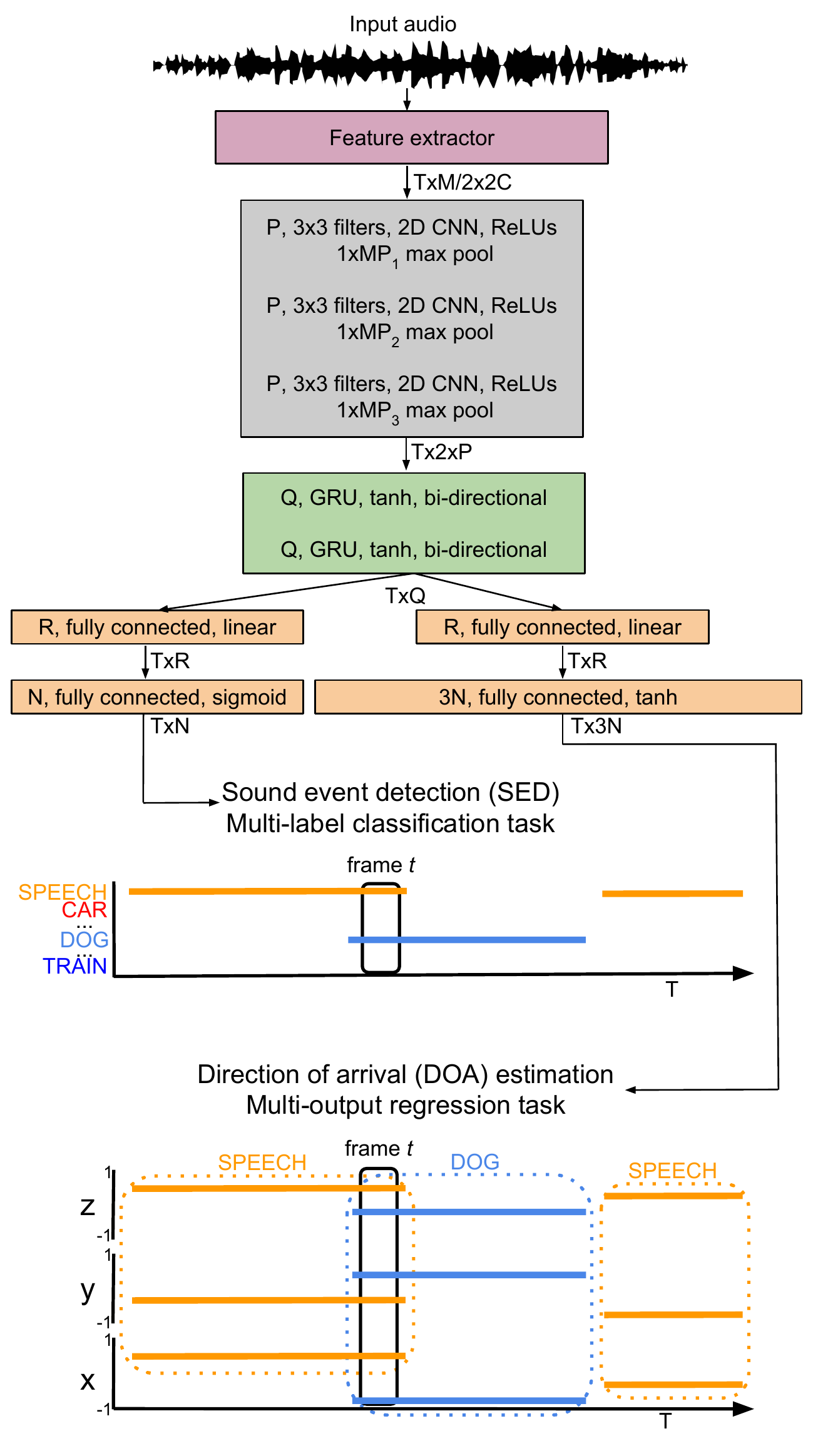} \label{fig:crnn}}}      
    \vspace{20pt}
    \subfloat[SELDnet output]{{\includegraphics[height=4.5cm, keepaspectratio, trim=0.3cm -0.2cm 0.5cm 0.1cm,clip]{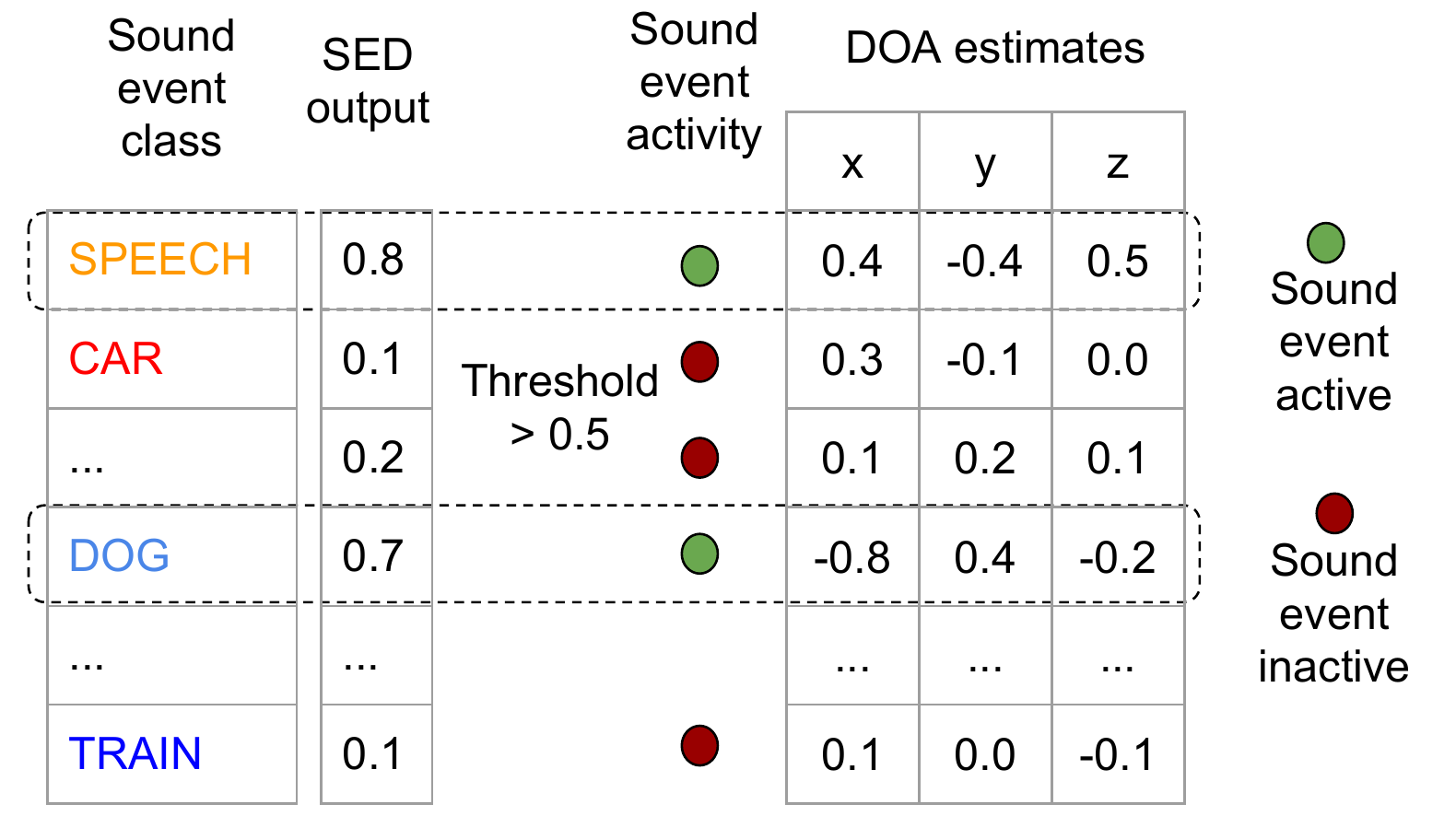} \label{fig:crnn_output}}}    
    \vspace{10pt}
  \caption{a) The proposed SELDnet and b) the frame-wise output for frame $t$ in Figure a). A sound event is said to be localized and detected when the confidence of the SED output exceeds the threshold.}
\end{figure}

The spectrogram is extracted from each of the $C$ channels of the multichannel audio using an $M$-point discrete Fourier transform (DFT) on Hamming window of length $M$ and 50\% overlap. The phase and magnitude of the spectrogram are then extracted and used as separate features. Only the $M/2$ positive frequencies without the zeroth bin are used. The output of the feature extraction block in Figure~\ref{fig:crnn} is a feature sequence of $T$ frames, with an overall dimension of $T\times M/2 \times 2C$, where the $2C$ dimension consists of $C$ magnitude and $C$ phase components.

\subsection{Neural network architecture}
The output of the feature extraction block is fed to the neural network as shown in Figure~\ref{fig:crnn}. In the proposed architecture the local shift-invariant features in the spectrogram are learned using multiple layers of 2D CNN. Each CNN layer has $P$ filters of $3\times3\times2C$ (as in~\cite{Adavanne2018_EUSIPCO}) dimensional receptive fields acting along the time-frequency-channel axis with a rectified linear unit (ReLU) activation. The use of filter kernels spanning all the channels allows the CNN to learn relevant inter-channel features required for localization, whereas the time and frequency dimensions of the kernel allows learning relevant intra-channel features suitable for both the DOA and SED tasks. After each layer of CNN, the output activations are normalized using batch normalization~\cite{batchNorm}, and the dimensionality is reduced using max-pooling ($MP_i$) along the frequency axis, thereby keeping the sequence length $T$ unchanged. The output after the final CNN layer with $P$ filters is of dimension $T\times2\times P$, where the reduced frequency dimension of $2$ is a result of max-pooling across CNN layers (see Section \ref{sssec:arch}). 

The output activation from CNN is further reshaped to a $T$ frame sequence of length $2P$ feature vectors and fed to bidirectional RNN layers which are used to learn the temporal context information from the CNN output activations.  Specifically, $Q$ nodes of gated recurrent units (GRU) are used in each layer with tanh activations. This is followed by two branches of FC layers in parallel, one each for SED and DOA estimation. The FC layers share weights across time steps. The first FC layer in both the branches contains $R$ nodes each with linear activation. The last FC layer in the SED branch consists of $N$ nodes with sigmoid activation, each corresponding to one of the $N$ sound event classes to be detected. The use of sigmoid activation enables multiple classes to be active simultaneously. The last FC layer in the DOA branch consists of $3N$ nodes with tanh activation, where each of the $N$ sound event classes is represented by $3$ nodes corresponding to the sound event location in $x$, $y$, and $z$, respectively. For a DOA estimate on a unit sphere centered at the origin, the range of location along each axes is $[-1, 1]$, thus we use the tanh activation for these regressors to keep the output of the network in a similar range. 

We refer to the above architecture as SELDnet. The SED output of the SELDnet is in the continuous range of $[0, 1]$ for each class, while the DOA output is in the continuous range of $[-1, 1]$ for each axes of the sound class location. A sound event is said to be active, and its respective DOA estimate is chosen if the SED output exceeds the threshold of 0.5 as shown in Figure~\ref{fig:crnn_output}. The network hyperparameters are optimized based on cross-validation as explained in Section~\ref{sssec:cv}.

\subsection{Training procedure}
In each frame, the target values for each of the active sound events in the SED branch output are one while the inactive events are zero. Similarly, for the DOA branch, the reference DOA $x$, $y$, and $z$ values are used as targets for the active sound events and $x=0$,  $y=0$, and $z=0$ is used for inactive events.  A binary cross-entropy loss is used between the SED predictions of SELDnet and reference sound class activities, while a mean square error (MSE) loss is used for the DOA estimates of the SELDnet and the reference DOA. By using the MSE loss for DOA estimation in 3D Cartesian coordinates we truly represent the distance between two points in space. The distance between two points $(x_1, y_1, z_1)$ and $(x_2, y_2, z_2)$ in 3D space is given by $\sqrt{SE}$, where $SE = (x_1-x_2)^2 + (y_1-y_2)^2 + (z_1-z_2)^2$, while the MSE between the same points is given by $SE/3$. Thus the MSE loss is simply a scaled version of the distance in 3D space, and reducing the MSE loss implies the reduction in the distance between the two points. 

Theoretically, the advantage of using Cartesian coordinates instead of azimuth and elevation for regression can be observed when predicting DOA in full azimuth and/or full elevation. The angles are discontinuous at the wrap-around boundary (for example the \ang{-180}, \ang{180} boundary for azimuth), while the Cartesian coordinates are continuous. This continuity allows the network to learn better. Further experiments on this are discussed in Section~\ref{ssec:experiments}.

We train the SELDnet with a weighted combination of MSE and binary cross-entropy loss for 1000 epochs using Adam optimizer with default parameters as used in the original paper~\cite{adamKeras}. Early stopping is used to control the network from over-fitting to training split. The training is stopped if the SELD score (Section~\ref{ssec:metric}) on the test split does not improve for 100 epochs. The network was implemented using Keras library~\cite{chollet2015keras} with TensorFlow~\cite{tensorflow2015-whitepaper} backend.

\section{Evaluation}
\label{sec:evaluation}
\subsection{Datasets}
The proposed SELDnet is evaluated on seven datasets that are summarized in Table~\ref{T:datasets}. Four of the datasets are synthesized with artificial impulse responses (IR), that consists of anechoic and reverberant scenarios virtually recorded both with a circular array and in the Ambisonics format. Three of the datasets are synthesized with real-life impulse responses, recorded with a spherical array and encoded into the Ambisonics format. All the datasets consist of stationary point sources each associated with a spatial coordinate. The synthesis procedure in all the datasets consists of mixing isolated sound event instances at different spatial locations, since this allows producing the reference event locations and times of activity for evaluation and training of the methods.

\subsubsection{TUT Sound Events 2018 - Ambisonic, Anechoic and Synthetic Impulse Response (ANSYN) dataset} This dataset consists of spatially located sound events in an anechoic environment synthesized using artificial IRs. It comprises three subsets: no temporally overlapping sources ($O1$), maximum two temporally overlapping sources ($O2$) and maximum three temporally overlapping sources ($O3$). Each of the subsets consists of three cross-validation splits with 240 training and 60 testing FOA format recordings of length 30 s sampled at 44100 Hz. The dataset is generated using the 11 isolated sound event classes from the DCASE 2016 task 2 dataset~\cite{dcase2016Task2} such as speech,
coughing, door slam, page-turning, phone ringing and keyboard. Each of these sound classes has 20 examples, of which 16 are randomly chosen for the training set and the rest four for the testing set, amounting to 176 examples from 11 classes for training, and 44 for testing. During synthesis of a recording, a random collection of examples are chosen from the respective set and are randomly placed in a spatial grid of \ang{10} resolution along azimuth and elevation, such that two overlapping sound events are separated by \ang{10}, and the elevation is in the range of $[\ang{-60}, \ang{60})$. In order to have a variability of amplitude, the sound events are randomly placed at a distance ranging from 1 to 10 m with 0.5 m resolution from the microphone. More details regarding the synthesis can be found in~\cite{Adavanne2018_EUSIPCO}.

\begin{table}[t]
\centering
\caption{Summary of datasets}
\label{T:datasets}
\resizebox{\columnwidth}{!}{
\begin{tabular}{l|ll|l|l}
Audio format & \multicolumn{1}{l|}{Sound scene} & \begin{tabular}[c]{@{}l@{}}Impulse\\response\end{tabular}   & \begin{tabular}[c]{@{}l@{}}Dataset\\acronym\end{tabular}  & \begin{tabular}[c]{@{}l@{}}Train/Test,\\notes\end{tabular} \\ \hline
\multirow{5}{*}{\begin{tabular}[c]{@{}l@{}}Ambisonic\\ (four channel)\end{tabular}} & \multicolumn{1}{l|}{Anechoic} & \multirow{2}{*}{Synthetic} & ANSYN & \multirow{3}{*}{240/60} \\ \cline{2-2} \cline{4-4}
 & \multicolumn{1}{l|}{\multirow{4}{*}{Reverberant}} &  & RESYN &  \\ \cline{3-4}
 & \multicolumn{1}{l|}{} & \multirow{3}{*}{Real life} & REAL &  \\ \cline{4-5} 
 & \multicolumn{1}{l|}{} &  & REALBIG & 600/150 \\ \cline{4-5} 
 & \multicolumn{1}{l|}{} &  & REALBIGAMB & \begin{tabular}[c]{@{}l@{}}600/150, \\ambiance\end{tabular} \\ \hline
\multirow{2}{*}{\begin{tabular}[c]{@{}l@{}}Circular array \\ (eight channel)\end{tabular}} & \multicolumn{1}{l|}{Anechoic} & \multirow{2}{*}{Synthetic} & CANSYN & \multirow{2}{*}{240/60} \\ \cline{2-2} \cline{4-4}
 & \multicolumn{1}{l|}{Reverberant} &  & CRESYN &  
\end{tabular}
}  

\end{table}

\subsubsection{TUT Sound Events 2018 - Ambisonic, Reverberant and Synthetic Impulse Response (RESYN) dataset} \label{sssec:resyn} This dataset is synthesized with the same details as the ambisonic ANSYN dataset, with the only difference being that the sound events are spatially placed within a room using the image source method~\cite{Allen1979}. Specifically, the microphone is placed at the center of the room, and the sound events are randomly placed around the microphone, with their distance ranging from 1 m from the microphone to the respective end of the room at 0.5 m resolution. The three cross-validation splits of each of the three subsets $O1$, $O2$ and $O3$ are generated for a moderately reverberant room of size $10\times 8\times 4$ m (Room 1), with reverberation times 1.0, 0.8, 0.7, 0.6, 0.5, and 0.4 s per each octave band, and 125 Hz--4 kHz band center frequencies. Additionally, to study the performance in mismatched reverberant scenarios, testing splits are generated for two different sized rooms: room 2 that is 80\% the volume ($8\times 8\times 4$ m) and reverberation-time per band of room 1, and room 3 that is 125\% the volume ($10\times 10\times 4$ m) and reverberation-time per band of room 1. In order to remove any ambiguity while comparing the performance difference of room 1 with room 2 and 3, we keep the sound events and their respective spatial locations in room 2 and 3 identical to the testing split of room 1.  But the individual sound events whose distance from the microphone exceeded the room size were reassigned a new distance within the room.  Further details on the reverberant synthesis can be read in~\cite{Adavanne2018_EUSIPCO}.

\subsubsection{TUT Sound Events 2018 - Ambisonic, Reverberant and Real-life Impulse Response (REAL) dataset}
\begin{figure}[t]
  \centering
  \centerline{\includegraphics[width=\linewidth, trim=0cm 0.1cm 0.1cm 0.1cm,clip]{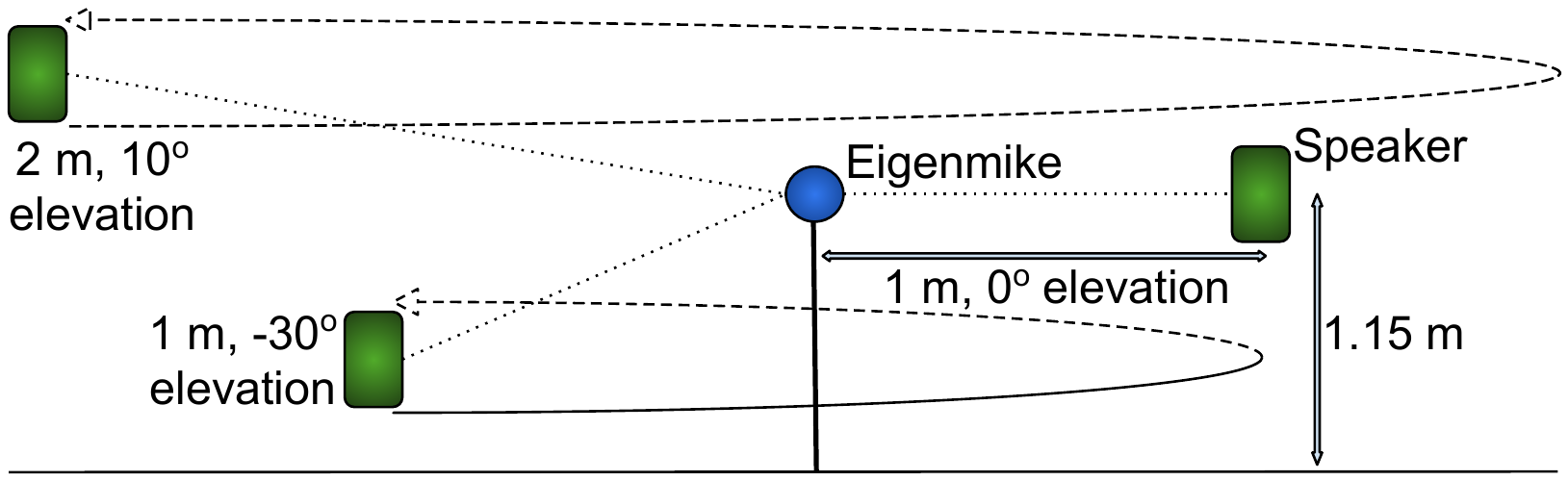}}
  \caption{Recording real-life impulse responses for sound scene generation. A person walks around the Eigenmike$^4$ holding a Genelec loudspeaker$^5$ playing a maximum length sequence at different elevation angles and distances.}
  \label{fig:seld_recording}
\end{figure}

In order to study the performance of SELDnet in a real-life scenario, we generated a dataset by collecting impulse responses from a real environment using the Eigenmike\footnote{\label{note1}https://mhacoustics.com/products} spherical microphone array. For the IR acquisition, we used a continuous measurement signal as in~\cite{enzner20093d}. The measurement was done by slowly moving a Genelec G Two loudspeaker\footnote{\label{note2}https://www.genelec.com/home-speakers/g-series-active-speakers} continuously playing a maximum length sequence around the array in circular trajectory in one elevation at a time, as shown in Figure~\ref{fig:seld_recording}. The playback volume was set to be 30 dB greater than the ambient sound level. The recording was done in a corridor inside the university with classrooms around it.

The moving-source IRs were obtained by a freely available tool from CHiME challenge~\cite{chime} which estimates the time-varying responses in STFT domain by forming a least-squares regression between the known measurement signal and the far-field recording independently at each frequency. The IR for any azimuth within one trajectory can be analyzed by assuming block-wise stationarity of acoustic channel. The average angular speed of the loudspeaker in the measurements was \ang{6}/s and we used a block size of 860 ms (81 STFT frames with analysis frame size of 1024 with 50 \% overlap and sample rate $F_s = 48$ kHz) for estimation of IR of length 170 ms (16 STFT frames).

The IRs were collected at elevations \ang{-40} to \ang{40} with \ang{10} increments at 1 m from the Eigenmike and at elevations \ang{-20} to \ang{20} with \ang{10} increments at 2 m. For the dataset creation, we analyzed the DOA of each time frame using MUSIC and extracted IRs for azimuthal angles at \ang{10} resolution (36 IRs for each elevation). The IR estimation tool~\cite{chime} was applied independently on all 32 channels of the Eigenmike. 

In order to synthesize the sound scene from the estimated IRs, we used isolated real-life sound events from the urbansound8k dataset~\cite{Salamon2014_acm}. This dataset consists of 10 sound event classes such as: air conditioner, car horn, children playing, dog barking, drilling, engine idling, gunshot, jackhammer, siren and street music. Among these, we did not include children playing and air conditioner classes since these can also occur in our ambiance recording which we use as background recording in dataset REALBIGAMB (Section~\ref{sssec:realbigamb}). From the sound examples in urbansound8k, we only used the ones marked as foreground in order to have clean isolated sound events. Similarly to the other datasets used in this paper, we used the splits 1, 8 and 9 provided in the urbansound8k as the three cross-validation splits. These splits were chosen as they had a good number of examples for all the chosen sound event classes after selecting only the foreground examples. The final selected examples varied in length from 100 ms to 4 s and amount to 15671.5 seconds from 4542 examples. 

During the sound scene synthesis, we randomly chose a sound event example and associated it with a random distance among the collected ones, azimuth and elevation angle. The sound event example was then convolved with the respective IR for the given distance, azimuth and elevation to spatially position it. Finally, after positioning all the sound events in a recording we converted this multichannel audio to FOA format. The transform of the microphone signals to FOA was performed using the tools published in \cite{politis_phd2016}. In total, we generated 300 such 30 s recordings in a similar fashion as ANSYN and RESYN with 240 of them earmarked for training and 60 for testing. Similar to the ANSYN recordings we also generated three subsets $O1$, $O2$ and $O3$ with a different number of overlapping sound events.

\subsubsection{TUT Sound Events 2018 - Ambisonic, Reverberant and Real-life Impulse Response big (REALBIG) dataset} 
In order to study the performance of SELDnet with respect to the size of the dataset, we generated for each of three ambisonic REAL subsets a 750 recordings REALBIG subset of 30 s length, with 600 for training and 150 for testing. 

\subsubsection{TUT Sound Events 2018 - Ambisonic, Reverberant, Real-life Impulse Response and Ambiance big (REALBIGAMB) dataset} \label{sssec:realbigamb}
Additionally, to simulate a real sound-scene we recorded 30 min of ambient sound to use as background noise in the same location as the IR recordings without changing the setup. We mixed randomly chosen segments of the recorded ambiance at three different SNRs: 0, 10 and 20 dB for each of the three ambisonic REALBIG subsets and refer to it as REALBIGAMB subsets. The ambiance used for the testing set was kept separate from the training set.

\subsubsection{TUT Sound Events 2018 - Circular array, Anechoic and Synthetic Impulse Response (CANSYN) dataset} 
To study the performance of SELDnet on generic array configurations, similarly to the SELD baseline method~\cite{Hirvonen2015} (Section~\ref{sssec:seldbaseline}), we synthesized the ANSYN recordings for a circular array of radius 5 cm with eight omnidirectional microphones at $0, 45, 90, 135, 180, 225, 270, 315^o$, and the array plane parallel to the ground, and refer to it as CANSYN. It is an exact replica of the ANSYN dataset in terms of the synthesized sound events except for the microphone array setup, and hence the number of channels. Similar to ANSYN, the CANSYN dataset has three subsets with a different number of overlapping sound events each with three cross-validation splits. 

\subsubsection{TUT Sound Events 2018 - Circular array, Reverberant and Synthetic Impulse Response (CRESYN) dataset} 
Similar to the CANSYN dataset, we synthesize the circular array version of ambisonic RESYN room 1 dataset, referred as CRESYN. During synthesis, the circular microphone array is placed in the center of the room, and the array plane parallel to the floor.

\subsection{Baseline methods}
The SELDnet is compared with six different baselines as summarized in Table~\ref{T:baseline}: two SED baselines (single- and multichannel), three DOA baselines (parametric and DNN-based), and a SELD baseline. 

\subsubsection{SED baseline}
The SED capabilities of the proposed SELDnet is compared with the existing state-of-the-art multichannel SED method~\cite{Adavanne2017}, referred here as MSEDnet. MSEDnet is easily scalable to any number of input audio channels and won~\cite{Adavanne_DCASE2017_binaural} the recently concluded real-life SED task in DCASE 2017~\cite{dcase2017}. In particular, it won the top two positions among 34 submissions, first using single-channel mode (referred as SEDnet) and a close second using multichannel mode. The SED performance of SELDnet is compared with both the single- and the multichannel modes of MSEDnet. 

In the original MSEDnet implementation~\cite{Adavanne2017} the input is a sequence of log mel-band energy (40-bands) frames, that are mapped to an equal-length sequence of sound event activities. The SED metrics (Section~\ref{ssec:metric}) for MSEDnet did not change much on using phase and magnitude components of the STFT spectrogram instead of log mel-band energies. Hence, in order to have a one-to-one comparison with SELDnet, we use the phase and magnitude components of the STFT spectrogram for MSEDnet in this paper. We train the MSEDnet for 500 epochs and use early stopping when SED score (Section~\ref{ssec:metric}) stops improving for 100 epochs.

\begin{table}[t]
\centering
\caption{Baseline and proposed method summary}

\label{T:baseline}
\resizebox{\columnwidth}{!}{
\begin{tabular}{l|l|l|l}
Task & Acronym & Notes & Datasets evaluated \\ \hline
\multirow{2}{*}{SED} & SEDnet~\cite{Adavanne2017} & Single channel & \multirow{2}{*}{All} \\
 & MSEDnet~\cite{Adavanne2017} & Multichannel &  \\ \hline
\multirow{3}{*}{DOA} & MUSIC\textsuperscript{*} & Azi and ele & \multirow{2}{*}{\begin{tabular}[c]{@{}l@{}}All except CANSYN \\ and CRESYN\end{tabular}} \\ 
 & DOAnet~\cite{Adavanne2018_EUSIPCO} & Azi and ele &  \\ \cline{4-4}
 & AZInet~\cite{Chakrabarty2017_nips} & Azi & \multirow{2}{*}{\begin{tabular}[c]{@{}l@{}}CANSYN and \\ CRESYN\end{tabular}} \\ \cline{1-3}
\multirow{3}{*}{SELD} & HIRnet~\cite{Hirvonen2015} & Azi &  \\ \cline{4-4}
 & SELDnet-azi & Azi & \multirow{2}{*}{All} \\
 & SELDnet & Azi and ele & \\
\multicolumn{4}{l}{\textsuperscript{*}\footnotesize{Parametric, all other methods are DNN based}}
\end{tabular}
}
\end{table}

\subsubsection{DOA baseline}
\label{music_baseline}
The DOA estimation performance of the SELDnet is evaluated with respect to three baselines. As a parametric baseline, we use MUSIC~\cite{Schmidt1986} and as DNN-based baselines, we use the recently proposed DOAnet~\cite{Adavanne2018_EUSIPCO} that estimates DOAs in 3D and~\cite{Chakrabarty2017_nips} that estimates only the DOA azimuth angle referred as AZInet. 

i) MUSIC: is a versatile high-resolution subspace method that can detect multiple narrowband source DOAs and can be applied to generic array setups. It is based on a subspace decomposition of the spatial covariance matrix of the multichannel spectrogram. For a broadband estimation of DOAs, we combine narrowband spatial covariance matrices over three frames and frequency bins from 50 to 8000 Hz. The steering vector information required to produce the MUSIC pseudo-spectrum from which the DOAs are extracted is adapted to the recording system under use, meaning uniform circular array steering vectors for CANSYN and CRESYN datasets, and real SH vectors for all the other ambisonic datasets.

MUSIC requires a good estimate of the number of active sound sources in order to estimate their DOAs. In this paper, we use MUSIC with the number of active sources taken from the reference of the dataset. Hence, the DOA estimation results of MUSIC can be considered as the best possible for the given dataset and serve as a benchmark for DOA estimation with and without the knowledge of the number of active sources. For a detailed description on MUSIC and other subspace methods, the reader is referred to \cite{Ottersten1993}, while for application of MUSIC to SH signals similar to this work, please refer to \cite{khaykin2012acoustic}.

ii) DOAnet: Among the recently proposed DNN-based DOA estimation methods listed in Table~\ref{T:doa_methods_summary}, the only method that attempts DOA estimation of multiple overlapping sources in 3D space is the DOAnet~\cite{Adavanne2018_EUSIPCO}. Hence, DOAnet serves as a suitable baseline to compare against the DOA estimation performance of the proposed SELDnet.
DOAnet is based on a similar CRNN architecture, the input to which is a sequence of multichannel phase and magnitude spectrum frames. It considers DOA estimation as a multi-label classification task by directional sampling with a resolution of \ang{10} along azimuth and elevation and estimating the likelihood of a sound source being active in each of these points.

iii) AZInet: Among the DOA-only estimation methods listed in Table~\ref{T:doa_methods_summary}, apart from the DOAnet~\cite{Adavanne2018_EUSIPCO}, methods~\cite{Chakrabarty2017_nips} and~\cite{He2018} are the only ones which attempt simultaneous DOA estimation of overlapping sources. Since~\cite{He2018} is evaluated on a dataset collected using microphones mounted on a humanoid robot, it is difficult to replicate the setup. Hence in this paper, we use the AZInet evaluated on a linear array in~\cite{Chakrabarty2017_nips} as the baseline. The AZInet is a CNN-based method that uses the phase component of the spectrogram of each channel as input, and maps it to azimuth angles in the range \ang{0} to \ang{180} at \ang{5} resolution as a multi-label classification task. AZInet uses only the phase spectrogram since the dataset evaluated on employs omnidirectional microphones, which for compact arrays and sources in the far-field, preserve directional information in inter-channel phase differences. Thus, although the evaluation in~\cite{Chakrabarty2017_nips} was carried out on a linear array, the method is generic to any omnidirectional array under these conditions. Further, in order to have a direct comparison, we extend the output of AZInet to full-azimuth with \ang{10} resolution and reduce the output of SELDnet to generate only the azimuth, i.e., we only estimate $x$ and $y$ coordinates of the DOA (SELDnet-azi). To enable this full-azimuth estimation we use the circular array with omnidirectional microphones datasets CANSYN and CRESYN.

\subsubsection{SELD baseline (HIRnet)}
\label{sssec:seldbaseline}
The joint SED and DOA estimation performance of SELDnet is compared with the method proposed by Hirvonen~\cite{Hirvonen2015}, hereafter referred to as HIRnet. The HIRnet was proposed for a circular array of omnidirectional microphones, hence we compare its performance only on the CANSYN and CRESYN datasets. HIRnet is a CNN-based network that uses the log-spectral power of each channel as the input feature and maps it to eight angles in full azimuth for each of the two classes (speech and music) as a multi-label classification task. More details about HIRnet can be found in~\cite{Hirvonen2015}. In order to have a direct comparison with SELDnet-azi, we extend HIRnet to estimate DOAs at a \ang{10} resolution for each of the sound event classes in our testing datasets.

\subsection{Evaluation metrics}
\label{ssec:metric}
The proposed SELDnet is evaluated using individual metrics for SED and DOA estimation. For SED, we use the standard polyphonic SED metrics, error rate (ER) and F-score calculated in segments of one second with no overlap as proposed in~\cite{metrics, CASSE:2018}. The segment-wise results are obtained from the frame-level predictions of the classifier by considering the sound events to be active in the entire segment if it is active in any of the frames within the segment. Similarly, we obtain labels for one-second segments of reference from its frame-wise annotation, and calculate the segment-wise ER and F-scores. Mathematically, the F-score is calculated as follows:
\begin{equation}
F = \frac{2 \cdot \sum_{k=1}^{K} TP(k)}{2 \cdot \sum_{k=1}^{K}TP(k)+ \sum_{k=1}^{K}FP(k)+ \sum_{k=1}^{K}FN(k)},
\label{Eqn:F}
\end{equation}
where the number of true positives $TP(k)$ is the total number of sound event classes that were active in both reference and predictions for the $k$th one-second segment. The number of false positives in a segment $FP(k)$ is the number of sound event classes that were active in the prediction but were inactive in the reference. Similarly, $FN(k)$ is the number of false negatives, i.e. the number of sound event classes inactive in the predictions but active in the reference.

The ER metric is calculated as
\begin{align}
ER = \frac{\sum_{k=1}^{K} S(k) + \sum_{k=1}^{K} D(k) + \sum_{k=1}^{K} I(k)}{\sum_{k=1}^{K} N(k)},
\label{Eqn:ER}
\end{align}
where, for each one-second segment $k$, $N(k)$ is the total number of active sound event classes in the reference. Substitution $S(k)$ is the number of times an event was detected but given the wrong level, this is obtained by merging the false negatives and false positives without individually correlating which false positive substitutes which false negative. The remaining false positives and false negatives, if any, are counted as insertions $I(k)$ and deletions $D(k)$ respectively. These statistics are mathematically defined as follows:
\begin{align}
& S(k) = \min(FN(k), FP(k)), \\
& D(k) = \max(0, FN(k)-FP(k)), \\
& I(k) = \max(0, FP(k)-FN(k)). 
\end{align}

An SED method is jointly evaluated using the F-score and ER metric, and an ideal method will have an F-score of one (reported as percentages in Table) and ER of zero. More details regarding the F-score and ER metric can be read in~\cite{metrics, CASSE:2018}.

The predicted DOA estimates ($x_E$, $y_E$, $z_E$) are evaluated with respect to the reference ($x_G$, $y_G$, $z_G$) used to synthesize the dataset, utilizing the central angle $\sigma\in[0,180]$. The $\sigma$ is the angle formed by ($x_E$, $y_E$, $z_E$) and ($x_G$, $y_G$, $z_G$) at the origin in degrees, and is given by
\begin{align}
\sigma = 2\cdot\arcsin\bigg(\frac{\sqrt{\Delta x^2 + \Delta y^2 + \Delta z^2}}{2}\bigg)\cdot\frac{180}{\pi},
\end{align}
where, $\Delta x = x_G - x_E$, $\Delta y = y_G - y_E$, and $\Delta z = z_G - z_E$. 
The DOA error for the entire dataset is then calculated as
\begin{align}
DOA\,error =  \frac{1}{D} \cdot \sum_{d=1}^{D} \sigma((x_G^d, y_G^d, z_G^d), (x_E^d, y_E^d, z_E^d)) 
\end{align}
where $D$ is the total number of DOA estimates across the entire dataset, and $\sigma((x_G^d, y_G^d, z_G^d), (x_E^d, y_E^d, z_E^d))$ is the angle between $d$-th estimated and reference DOAs. 

Additionally, in order to account for time frames where the number of estimated and reference DOAs are unequal, we report the frame recall, calculated as $TP / (TP +  FN)$ in percentage, where true positives $TP$ is the total number of time frames in which the number of DOAs predicted is equal to reference, and false negatives $FN$ is the total number of frames where the predicted and reference DOA are unequal.

The DOA estimation method is jointly evaluated using the DOA error and the frame recall, and an ideal method has a frame recall of one (reported as percentages in Table) and DOA error of zero.

During the training of SELDnet, we perform early stopping based on the combined SELD score calculated as
\begin{equation}
SELD\,score = (SED\,score + DOA\,score)/2,
\end{equation}
where
\small
\begin{equation}
SED\,score = (ER + (1-F))/2, \\
\end{equation}
\begin{equation}
DOA\,score = \big(DOA\,error / 180 + (1 - frame\,recall)\big)/2,
\end{equation}
\normalsize
and an ideal SELD method will have an SELD score of zero. In the proposed method, the localization performance is dependent on the detection performance. This relation is represented by the frame recall metric of DOA estimation. As a consequence, the SELD score which is comprised of frame recall metric in addition to the SED metrics can be seen to weigh the SED performance more than DOA.

\subsection{Experiments}
\label{ssec:experiments}
The SELDnet is evaluated in different dimensions to understand its potential and drawbacks. The experiments carried out with different datasets in this regard are explained below.

\subsubsection{SELDnet architecture and model parameter tuning} \label{sssec:cv}
A wide variety of architectures with different combinations of CNN, RNN and FC layers are explored on the ANSYN $O2$ subset with frame length $M=1024$ ($23.2\, ms$). Additionally, for each architecture, we tune the model parameters such as the number of CNN, RNN, and FC layers (0 to 4) and nodes (in the set of $[16, 32, 64, 128, 256, 512]$). The input sequence length is tuned in the set of $[32, 64, 128, 256, 512]$, the DOA and SED branch output loss weights in the set of $[1, 5, 50, 500]$, the regularization (dropout in the set of $[0, 0.1, 0.2, 0.3, 0.4, 0.5]$, L1 and L2 in the set of $[0, \SI{e-1}, \SI{e-2}, \SI{e-3}, \SI{e-4}, \SI{e-5}, \SI{e-6}, \SI{e-7}]$) and the CNN max-pooling in the set of $[2, 4, 6, 8, 16]$ for each layer. The best set of parameters are the ones which give the lowest SELD score on the three cross-validation splits of the dataset. After finding the best network architecture and configuration, we tune the input audio feature parameter $M$ by varying it in the set of $[512, 1024, 2048]$. Simultaneously the sequence length is also changed with respect to $M$ such that the input audio length is kept constant ($1.49\,s$ obtained from the first round of tuning). We perform fine-tuning of model parameters for different $M$ and sequence length values, this time only the number of CNN, RNN and FC nodes are tuned in a small range (neighboring nodes in the set of [16, 32, 64, 128, 256, 512]) to identify the optimum parameters. Similar fine-tuning is repeated for other datasets.

\subsubsection{Selecting SELDnet output format}
The output format for polyphonic SED in the literature has become standardized to estimating the temporal activity of each sound class using frame-wise binary numbers~\cite{Hayashi_TASLP2017,Zohrer_INTERSPEECH2017,Zhang2015,Phan2016}. On the other hand, the output formats for DOA estimation are still being experimented with as seen in Table~\ref{T:doa_methods_summary}. Among the DOA estimation methods using regression mode, there are two possible output formats, predicting azimuth and elevation, and predicting $x,y,z$ coordinates of the DOA on the unit sphere. In order to identify the best output format among these two, we evaluate the SELDnet for both. During this evaluation, only the output weight parameter of the model is fine-tuned in the set of $[1, 5, 50, 500]$. Additionally, for a regression-based model, the default output i.e. the DOA target when the event is not active should be chosen carefully. In this study, we chose the default DOA output to be \ang{180} in azimuth and \ang{60} in elevation (the datasets do not contain sound events for these DOA values), and $x=0$, $y=0$ and $z=0$ for default Cartesian outputs. The chosen default Cartesian coordinates are equidistant from all the possible DOA values. On the other hand, there are no such equidistant azimuth and elevation values. Hence we chose the default values (\ang{180}, \ang{60}) to be in a similar range as the true DOA values.

\subsubsection{Continuous DOA estimation and performance on unseen DOA values} \label{sssec:cont_doa}
Theoretically, the advantage of using a regression-based DOA estimator over a classification-based one is that the network is not limited to a set of DOA angles, but it can operate as a high-resolution continuous DOA estimator. To study this, we train the SELDnet on ANSYN subsets whose sound events are placed on an angular grid of \ang{10} resolution along azimuth and elevation, and test the model on a dataset where the angular grid is shifted by \ang{5} along azimuth and elevation while keeping the temporal location unchanged. This shift makes the DOA values of the testing split unseen, and correctly predicting the DOAs will prove that the regression model can estimate the DOAs in a continuous space. Additionally, it also proves the robustness of the SELDnet to predict unseen DOA values.

\subsubsection{Performance on mismatched reverberant dataset}
Parametric DOA estimation methods are known to be sensitive to reverberation~\cite{dibiase2001robust}. In this regard, we first evaluate the performance of SELDnet on the simulated (RESYN), and real-life (REAL, REALBIG, and REALBIGAMB) reverberant datasets and further compare the results with the parametric baseline MUSIC.

DNN based methods are known to fail when the training and testing splits come from different domains. For example, the performance of a DNN trained on anechoic dataset would be poor on a reverberant testing dataset. This performance can only be improved by training the DNN on a similar reverberant dataset as the testing dataset. On the other hand, it is impractical to train such a DNN for every existing room-dimension, its surface material distribution, and the reverberation times associated with it. In this regard, it would be ideal if the proposed method is robust to a moderate mismatch in reverberant conditions so that a single model can be used for a range of comparable room configurations. Motivated by this, we study the sensitivity of SELDnet on moderately mismatched reverberant data. Specifically, we train the SELDnet with RESYN room 1 dataset and test it on RESYN room 2 and 3 datasets that are mismatched in terms of volume and reverberation times as described in Section~\ref{sssec:resyn}.

\subsubsection{Performance on the size of the dataset}
We study the performance of SELDnet on two datasets, REAL, and REALBIG that are similar in content, but different in size.

\subsubsection{Performance with ambiance at different SNR}

The performance of SELDnet with respect to different SNRs (0, 10 and 20 dB) of the sound event is studied on the REALBIGAMB dataset.

\begin{figure*}[!htb]
\centering	
    \minipage{0.5\linewidth}
    \includegraphics[height=4cm, width=\linewidth, trim=0.25cm 0.25cm 0.25cm 0.25cm,clip, keepaspectratio]{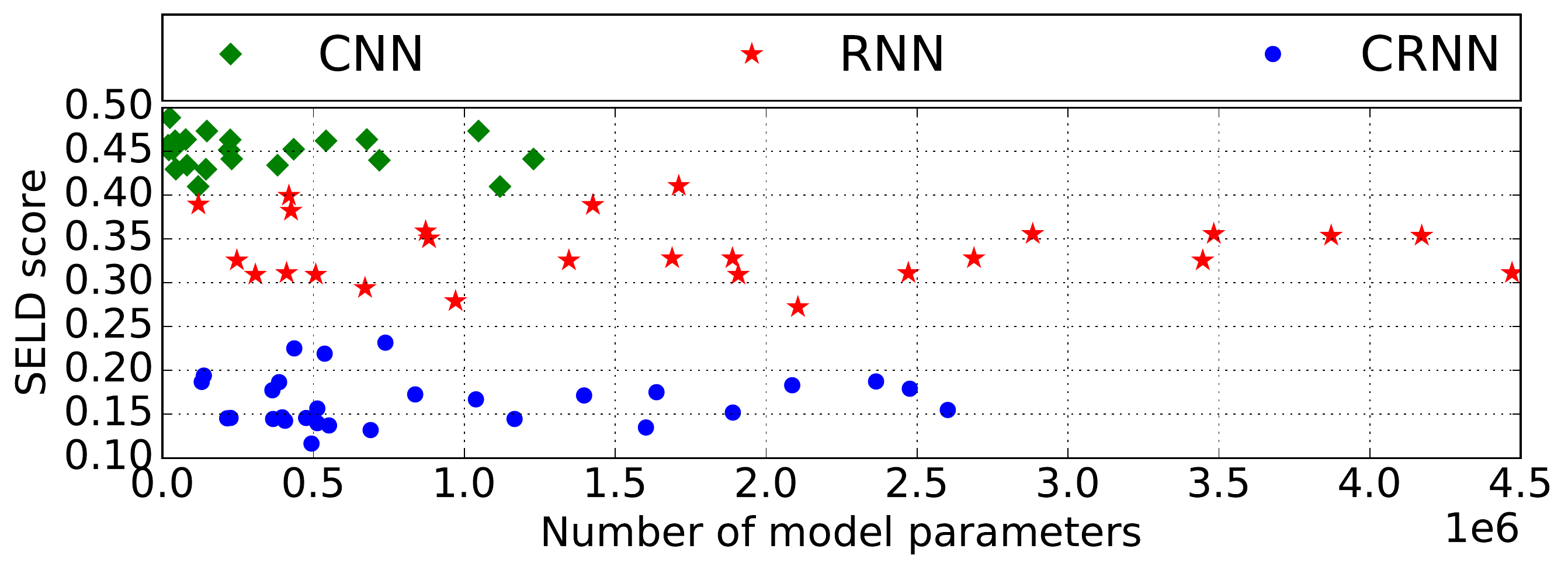}
    \endminipage\hfill
    \minipage{0.5\linewidth}
    \includegraphics[height=4cm, width=\linewidth, trim=0.25cm 0.25cm 0.25cm 0.25cm,clip, keepaspectratio]{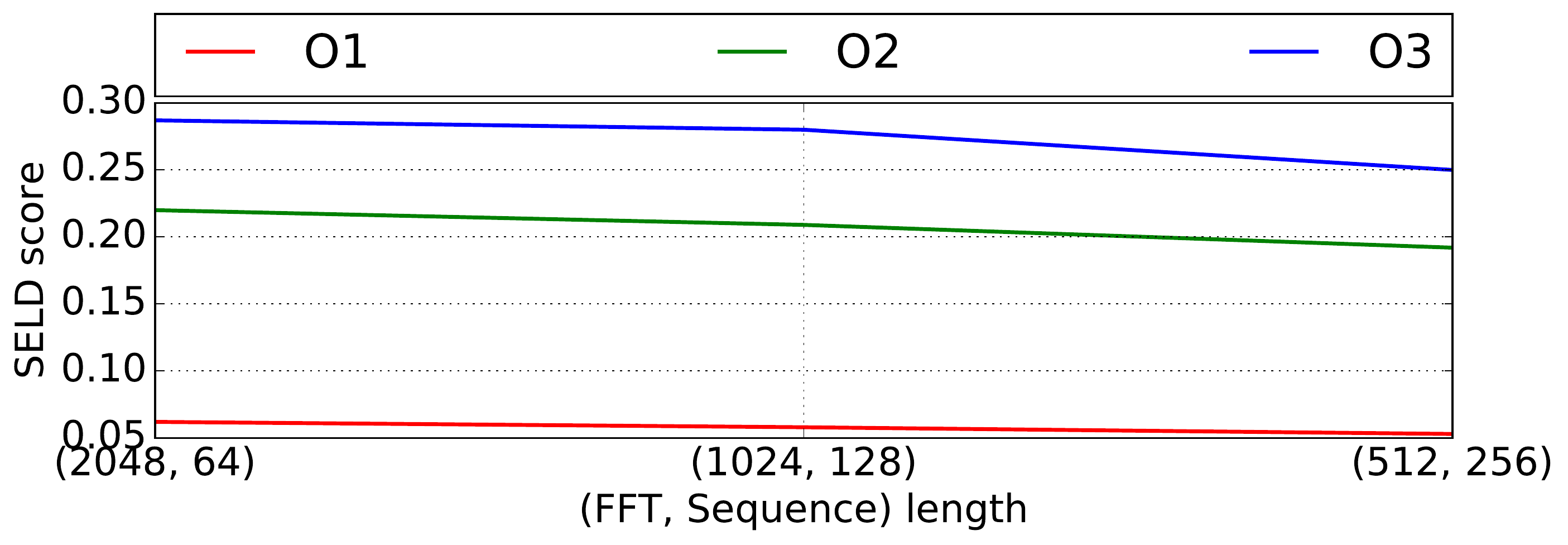}
    \endminipage\hfill
    
    \hspace{-15pt}
    \minipage{0.45\linewidth}
    \caption{SELD score for ANSYN $O2$ dataset for different CNN, RNN and CRNN architecture configurations.}
    \label{fig:modelparams}
    \endminipage \hspace{15pt}
    \minipage{0.45\linewidth}
    \caption{SELD score for ANSYN datasets for different combinations of FFT length and input sequence length in frames.}
    \label{fig:fft_seq_seld}
    \endminipage\hfill
    
    \vspace{20pt}
    \minipage{0.5\linewidth}
    \includegraphics[height=4cm, width=\linewidth, trim=0.25cm 0.25cm 0.25cm 0.25cm,clip, keepaspectratio]{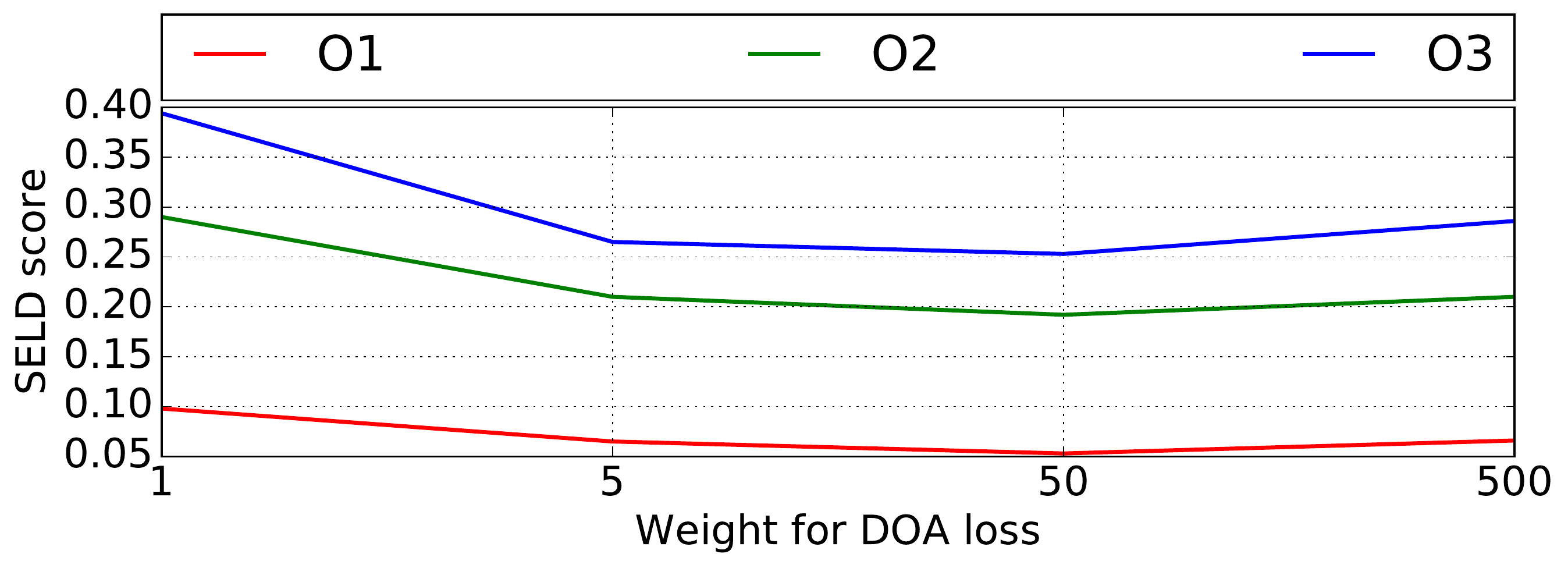}
    \endminipage\hfill
    \minipage{0.5\linewidth}
    \includegraphics[height=4cm, width=\linewidth, trim=0.25cm 0.25cm 0.25cm 0.25cm,clip,keepaspectratio]{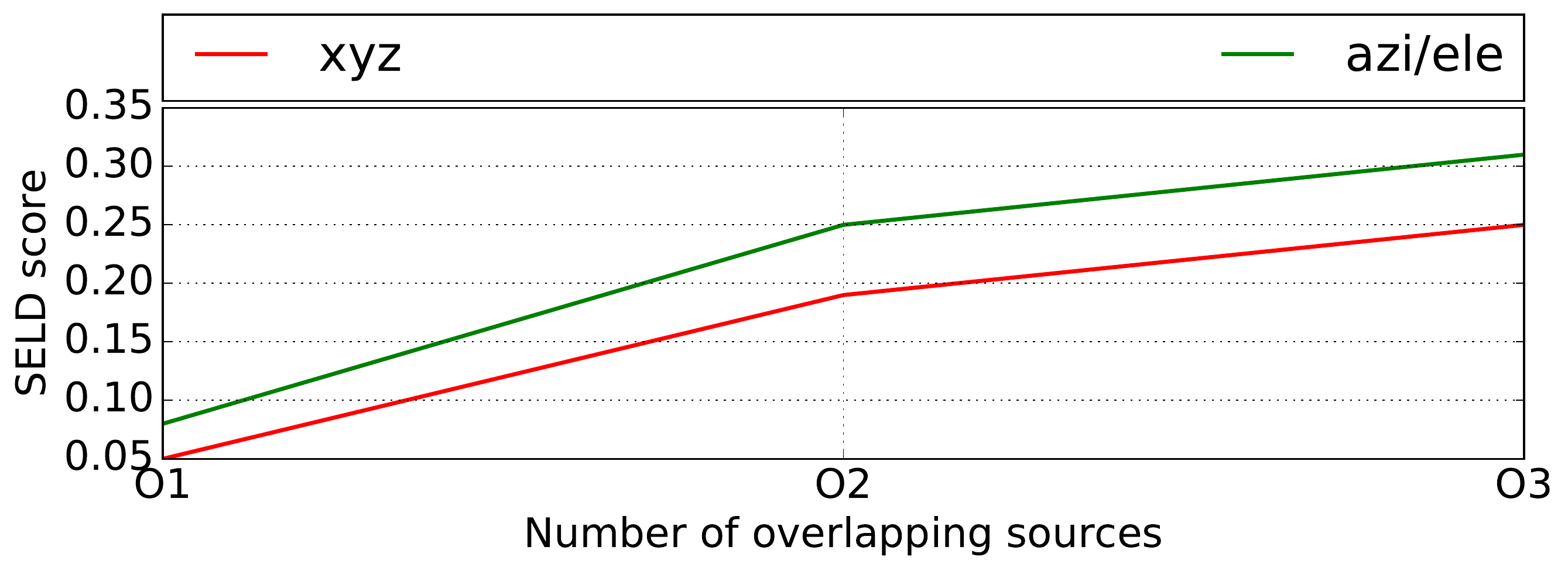}
    \endminipage\hfill
    
    \hspace{-15pt}
    \minipage{0.45\linewidth}
    \caption{SELD score for ANSYN datasets with respect to different weights for DOA output.}
    \label{fig:doa_wt}
    \endminipage \hspace{15pt}
    \minipage{0.45\linewidth}
    \caption{SELD score for ANSYN datasets with respect to DOA output formats.} 
    \label{fig:xyz_azi_ele}
    \endminipage \hfill
\end{figure*}

\subsubsection{Generic to array structure}
SELDnet is a generic method that learns to localize and recognize sound events from any array structure. This additionally implies that the SELDnet will continue to work in the desired manner if the configuration of the array such as individual microphone response, microphone spacing and the number of microphones remains the same between the training and testing set. If the array configuration changes between the training and testing set, then the SELDnet will have to be retrained for the new array configuration.

In order to prove that the SELDnet is applicable to any array configuration and not just dependent on the Ambisonics format, SELDnet is evaluated on a circular array. In comparison to the Ambisonic format, the chosen circular array has a different number of microphones,  each placed on a single plane, and with an omnidirectional response. Further, we compare the SELDnet performance with dataset compatible baselines such as SEDnet, MSEDnet, HIRnet, and AZInet. Since the HIRnet and AZInet baselines methods are proposed for estimating azimuth only, we compare the results with the SELDnet-azi version. Additionally, we also report the performance of using SELDnet with DOA estimation in $x, y, z$ axis on CANSYN and CRESYN datasets.

In general, for all our experiments the only difference between the training and testing splits is the mutually exclusive set of sound examples. Apart from experiment~\ref{sssec:cont_doa} the training and testing splits contains the same set of spatial locations i.e., azimuth and elevation angles at \ang{10} resolution amounting to 468 spatial locations (= 36 azimuth angles * 13 elevation angles). But the distance of the sound event at each of this 468 spatial locations is an added variable. For example, in the anechoic case, a sound event can be placed anywhere between 1-10 m at 0.5 m resolution. This variable amounts to 8892 spatial locations (= 468 * 19 distance positions) that are being coarsely grouped to 468 locations. This complexity is stretched further in experiment~\ref{sssec:cont_doa} where the testing split sound event examples and their spatial locations both are different from the training split.

\section{Results and Discussion}
\label{sec:results}
\subsubsection{SELDnet architecture and model parameter tuning}
\label{sssec:arch}

The SELD scores obtained with hyper-parameter tuning of different CNN, RNN, and CRNN configurations as explained in Section~\ref{sssec:cv} are visualized with respect to the number of model parameters in Figure~\ref{fig:modelparams}. CNN in this figure refers to a SELDnet architecture which had no RNN layers but just CNN and FC layers. Similarly, RNN refers to SELDnet without CNN layers, while CRNN refers to SELDnet with CNN, RNN and FC layers. This experiment was carried out on ANSYN $O2$ dataset. The CRNN architecture was seen to perform the best followed by the RNN architecture.

The optimum model parameters across the ANSYN subsets after hyper-parameter tuning the CRNN architecture was found to have three layers of CNN with 64 nodes each ($P$ in Figure~\ref{fig:crnn}), followed by two layers of GRU with 128 nodes each ($Q$ in Figure~\ref{fig:crnn}), and one FC layer with 128 nodes ($R$ in Figure~\ref{fig:crnn}). The max-pooling over frequency after each of the three CNN layers ($MP_i$ in Figure~\ref{fig:crnn}) was $(8, 8, 2)$. This configuration had about 513,000 parameters.


\begin{figure*}[!b]
  \centering
  \subfloat[ANSYN $O1$]{{\includegraphics[width=0.5\linewidth, trim=0.35cm 0.4cm 0.35cm 0.3cm,clip]{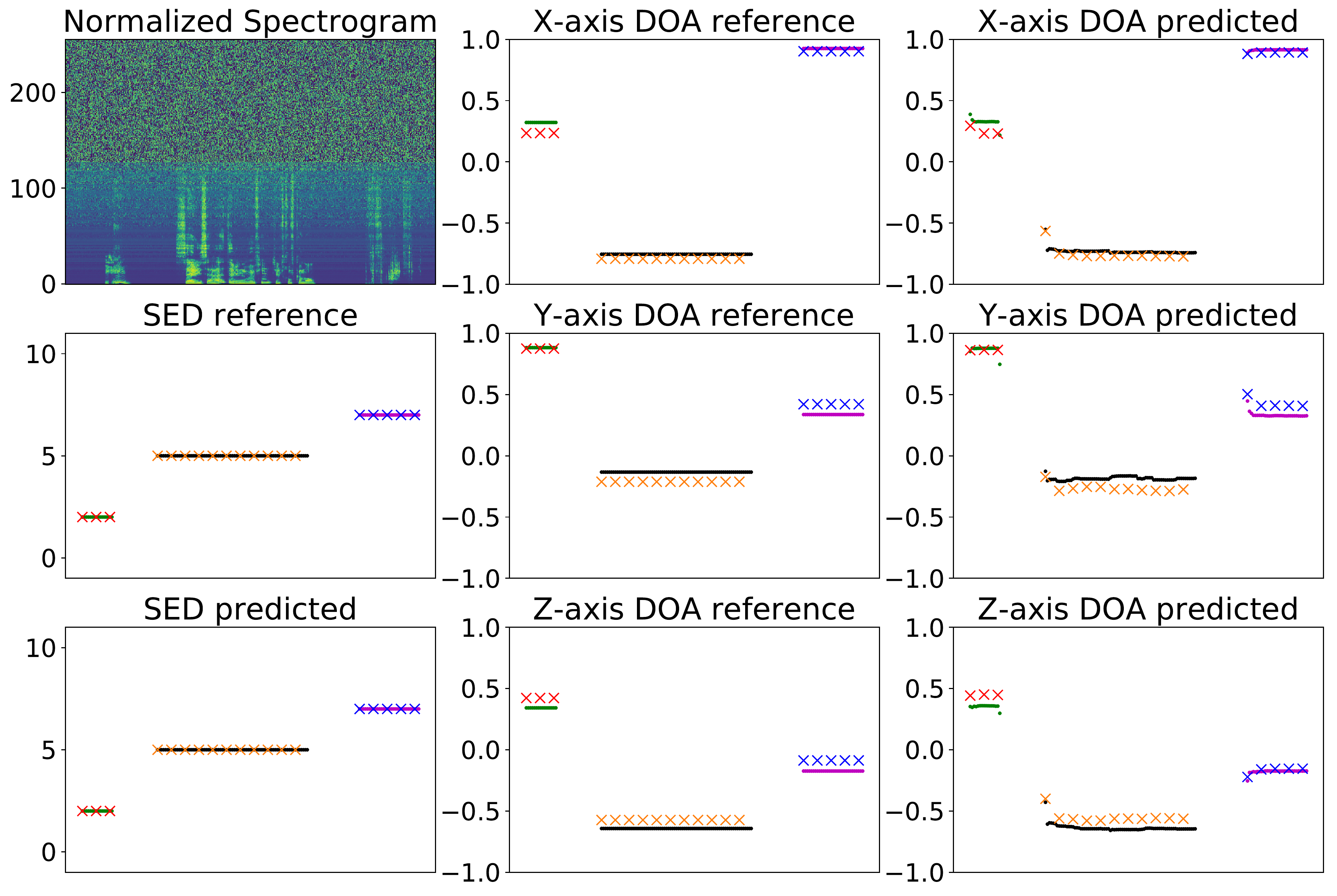} \label{fig:ov1_results}}}  
  \subfloat[ANSYN $O2$]{{\includegraphics[width=0.5\linewidth, trim=0.35cm 0.4cm 0.35cm 0.3cm,clip]{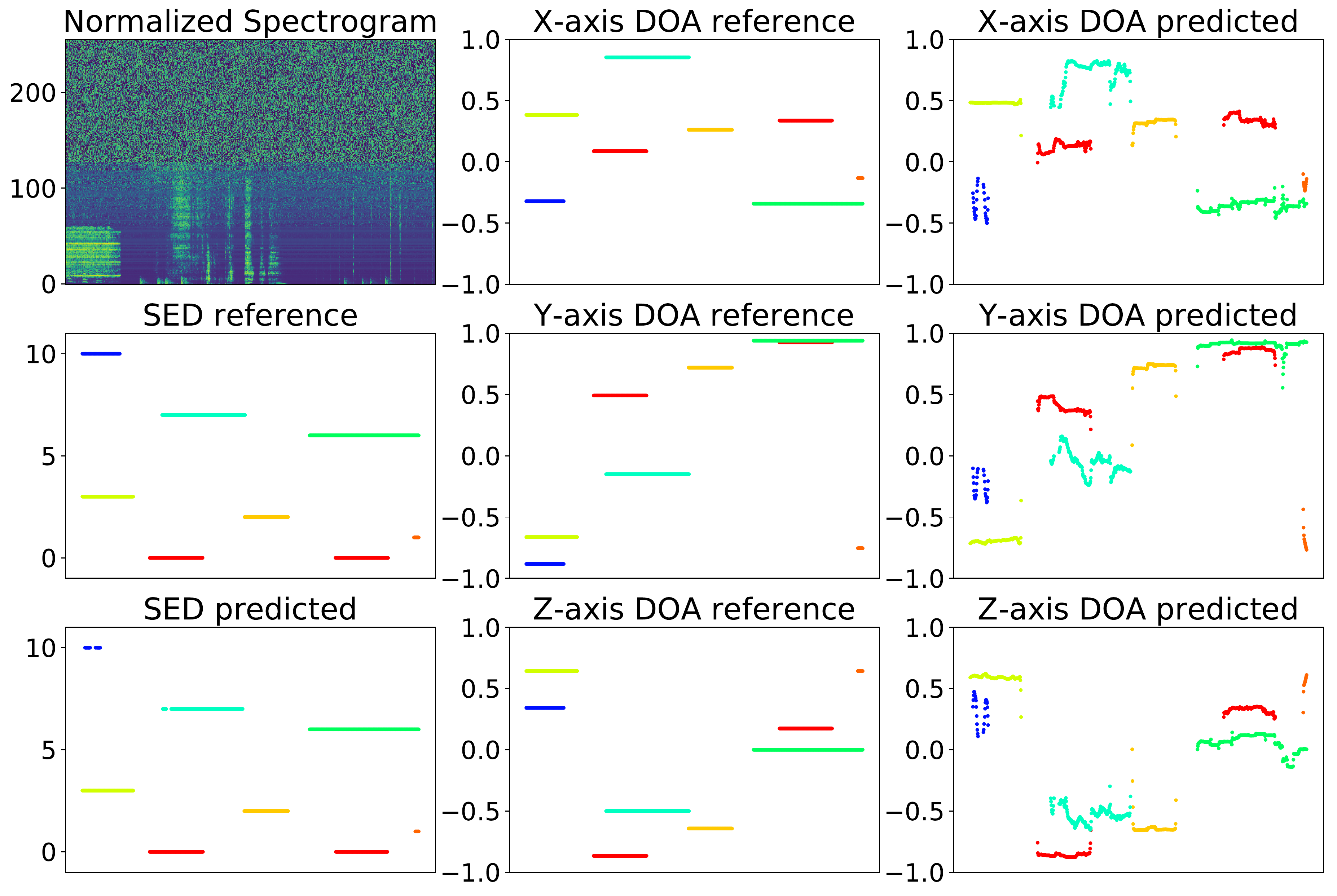} \label{fig:ov2_results}}}
\caption{SELDnet input and outputs visualized for ANSYN $O1$ and $O2$ datasets. The horizontal-axis of all sub-plots for a given dataset represents the same time frames, the vertical-axis for spectrogram sub-plot represents the frequency bins, vertical-axis for SED reference and prediction sub-plots represents the unique sound event class identifier, and for the DOA reference and prediction sub-plots, it represents the distance from the origin along the respective axes. The bold lines visualize both the reference labels and predictions of DOA and SED for ANSYN $O1$ and $O2$ datasets, while the \texttimes{} markers in Figure~\ref{fig:ov1_results} visualize the results for testing split with unseen DOA values (shifted by \ang{5} along azimuth and elevation).}
\label{fig:crnn_op_visual}
\end{figure*}

\begin{table*}[!htb]
\centering
\caption{SED and DOA estimation metrics for ANSYN and RESYN datasets. The results for the RESYN Room 2 and 3 testing splits were obtained from classifiers trained on RESYN Room 1 training set. Best scores for subsets in bold.}
\label{T:ansim_resim_results}
\begin{tabular}{l|l|lll|lll|lll|lll|}
 \multicolumn{2}{l|}{} & \multicolumn{3}{c|}{ANSYN} & \multicolumn{3}{c|}{RESYN Room 1}  & \multicolumn{3}{c|}{RESYN Room 2}  & \multicolumn{3}{c|}{RESYN Room 3} \\ \cline {3-14}
 \multicolumn{1}{l|}{}& Overlap & 1 & 2 & 3 & 1 & 2 & 3 & 1 & 2 & 3 & 1 & 2 & 3 \\ \cline {2-14}
\multicolumn{14}{l}{SED metrics} \\\hline
SELDnet & ER &\bf 0.04 & 0.16 & 0.19 &\bf 0.10 & 0.29 & 0.32 & \bf0.11 & 0.33 & 0.35 &\bf 0.13 & 0.32 & 0.34 \\
 & F & \bf97.7 & 89.0 & 85.6 &\bf 92.5 & \bf79.6 &\bf 76.5 &\bf 91.6 & \bf79.5 & \bf75.8 & \bf89.8 & 79.1 & 75.5 \\ \cline{2-14}
MSEDnet~\cite{Adavanne2017} & ER & 0.10 & \bf0.13 & \bf0.17 & 0.17 &\bf 0.28 & \bf0.29 & 0.19 & \bf0.30 & \bf0.26 & 0.18 & \bf0.29 & \bf0.30 \\
 & F & 94.4 &\bf 90.1 & \bf87.2 & 89.1 & 79.1 & 75.6 & 88.3 & 78.2 & 74.2 & 86.5 & \bf80.5 &\bf 76.1 \\ \cline{2-14}
SEDnet~\cite{Adavanne2017} & ER & 0.14 & 0.16 & 0.18 & 0.18 & \bf0.28 & 0.30 & 0.19 & 0.32 & 0.28 & 0.21 & 0.32 & 0.33 \\
 & F & 91.9 & 89.1 & 86.7 & 88.2 & 76.9 & 74.1 & 87.6 & 76.4 & 73.2 & 85.1 & 78.2 & 75.6 \\
 
\multicolumn{14}{l}{DOA metrics} \\ \hline
SELDnet & DOA error & 3.4 & 13.8 & 17.3 & 9.2 & 20.2 & \bf26.0 & 11.5 & 26.0 & \bf33.1 & 12.1 & 25.4 & \bf31.9 \\
 & Frame recall & \bf99.4 & \bf85.6 & \bf70.2 & \bf95.8 & \bf74.9 & \bf56.4 & \bf96.2 & \bf78.9 & \bf61.2 & \bf95.9 & \bf78.2 & \bf60.7 \\\cline{2-14}
DOAnet~\cite{Adavanne2018_EUSIPCO} & DOA error & \bf0.6 & 8.0 & 18.3 & \bf6.3 &\bf 11.5 & 38.4 & \bf3.4 & \bf6.9 & - & \bf4.6 & \bf10.9 & - \\
 & Frame recall & 95.4 & 42.7 & 1.8 & 59.3 & 15.8 & 1.2 & 46.2 & 14.3 & - & 49.7 & 14.1 & - \\\cline{2-14}
 
MUSIC & DOA error & 4.1 & \bf7.2 & \bf15.8 & 40.2 & 47.1 & 50.5 & 45.7 & 58.1 & 74.0 & 48.3 & 60.6 & 75.6
\end{tabular}

\end{table*}


Further, the SELDnet was seen to perform best with no regularization (dropout, or L1 or L2 regularization of weights). A frame length of $M=512$ and sequence length of 256 frames was seen to give the best results across ANSYN subsets (Figure~\ref{fig:fft_seq_seld}). Furthermore, on tuning the sequence length with frame length fixed ($M=512$), the best scores were obtained using 512 frames ($2.97\,s$). Sequences longer than this could not be studied due to hardware restrictions. For the output weights, DOA output weighted 50 times more than SED output was seen to give the best results across subsets (Figure~\ref{fig:doa_wt}).

On fine-tuning the SELDnet parameters obtained with ANSYN dataset for RESYN subsets, the only parameter that helped improve the performance was using a sequence length of 256 instead of 512, leaving the total number of network parameters unchanged at 513,000. Similar configuration gave the best results for CANSYN and CRESYN datasets.

Model parameters identical to ANSYN dataset were observed to perform the best on the REAL subsets. The same parameters were also used for the study of REALBIG and REALBIGAMB subsets.


\begin{figure}[!b]
\captionsetup[subfigure]{width=6em}
\centering		
    \subfloat[ANSYN $O1$]{{\includegraphics[width=0.45\linewidth, height=2.1cm, keepaspectratio, trim=0cm 0cm 0.5cm 0.5cm,clip]{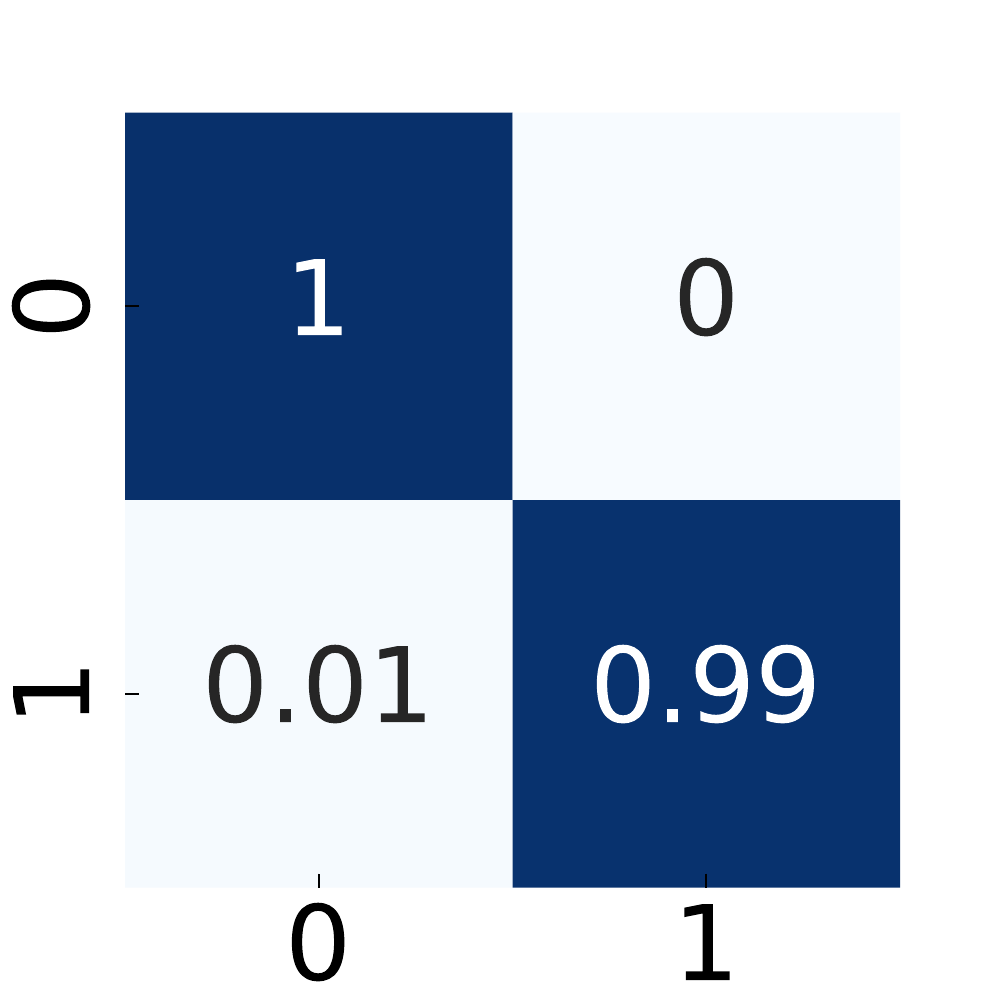}}}%
    \hspace{15pt}     
    \subfloat[RESYN $O1$]{{\includegraphics[width=0.45\linewidth, height=2.1cm, keepaspectratio, trim=0cm 0cm 0.5cm 0.5cm,clip]{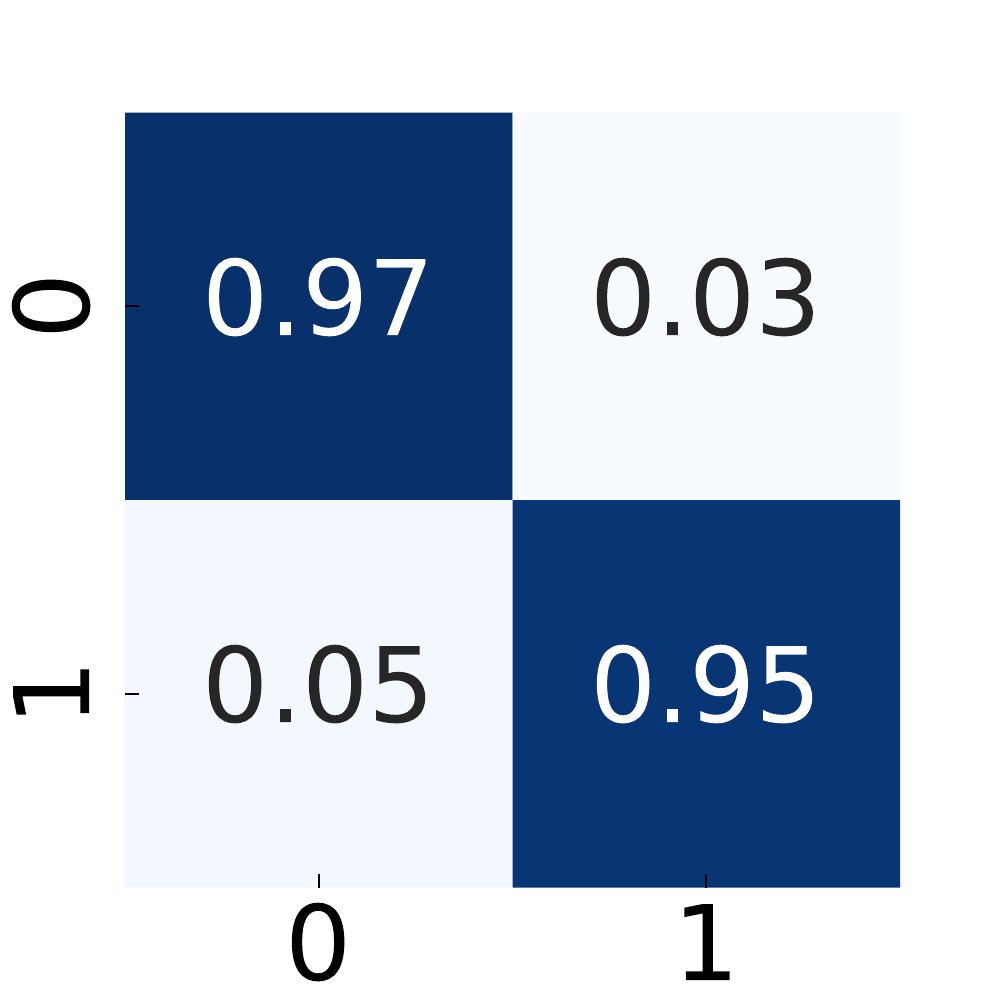}}}%
    \vspace{3pt}
    \subfloat[ANSYN $O2$]{{\includegraphics[width=0.45\linewidth, height=3cm, keepaspectratio, trim=1cm 0.75cm .25cm 2cm,clip]{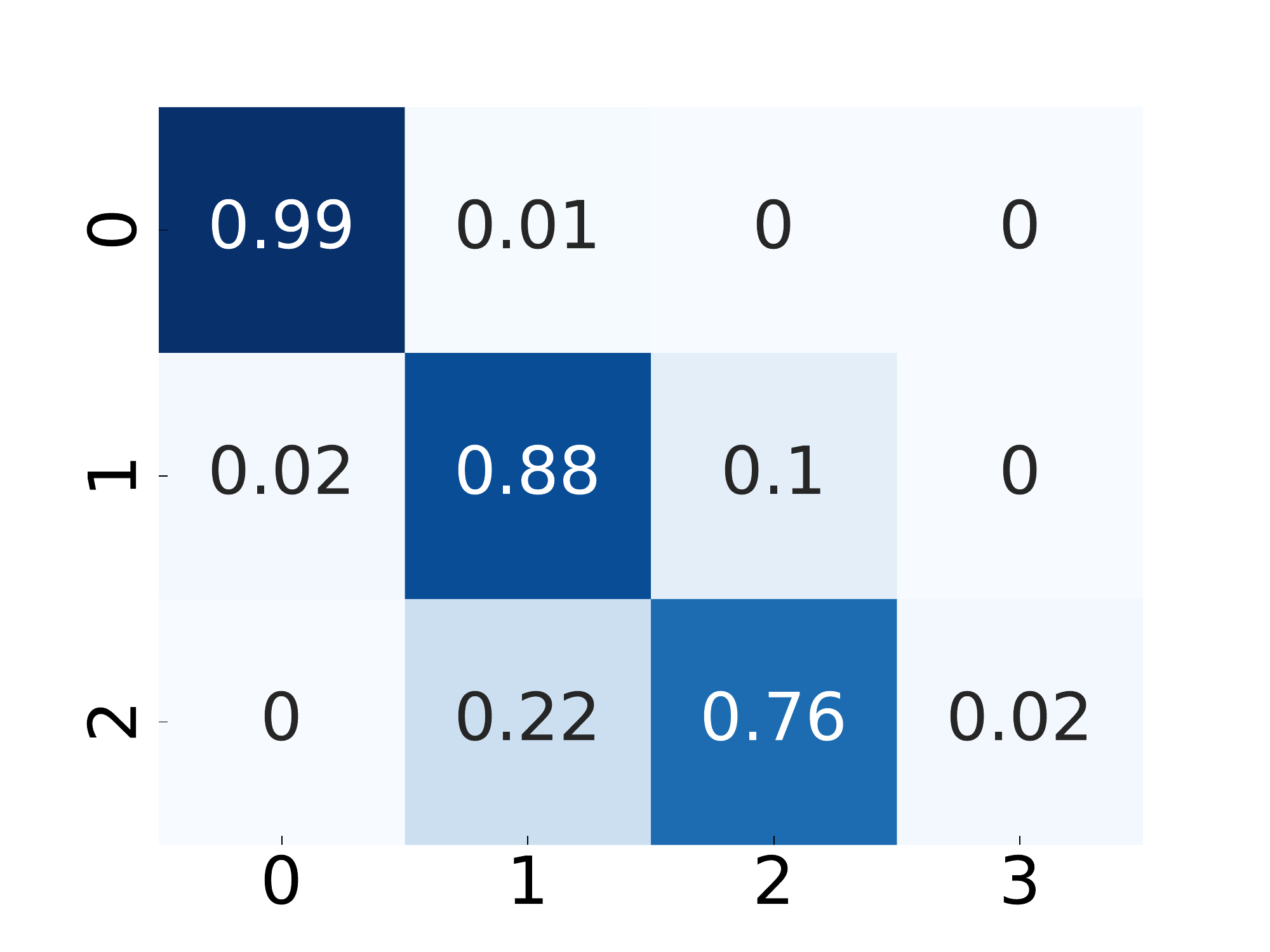}}\label{fig:ansyn_o2}}%
    \subfloat[RESYN $O2$]{{\includegraphics[width=0.45\linewidth, height=3cm, keepaspectratio, trim=1cm 0.75cm 0cm 2cm,clip]{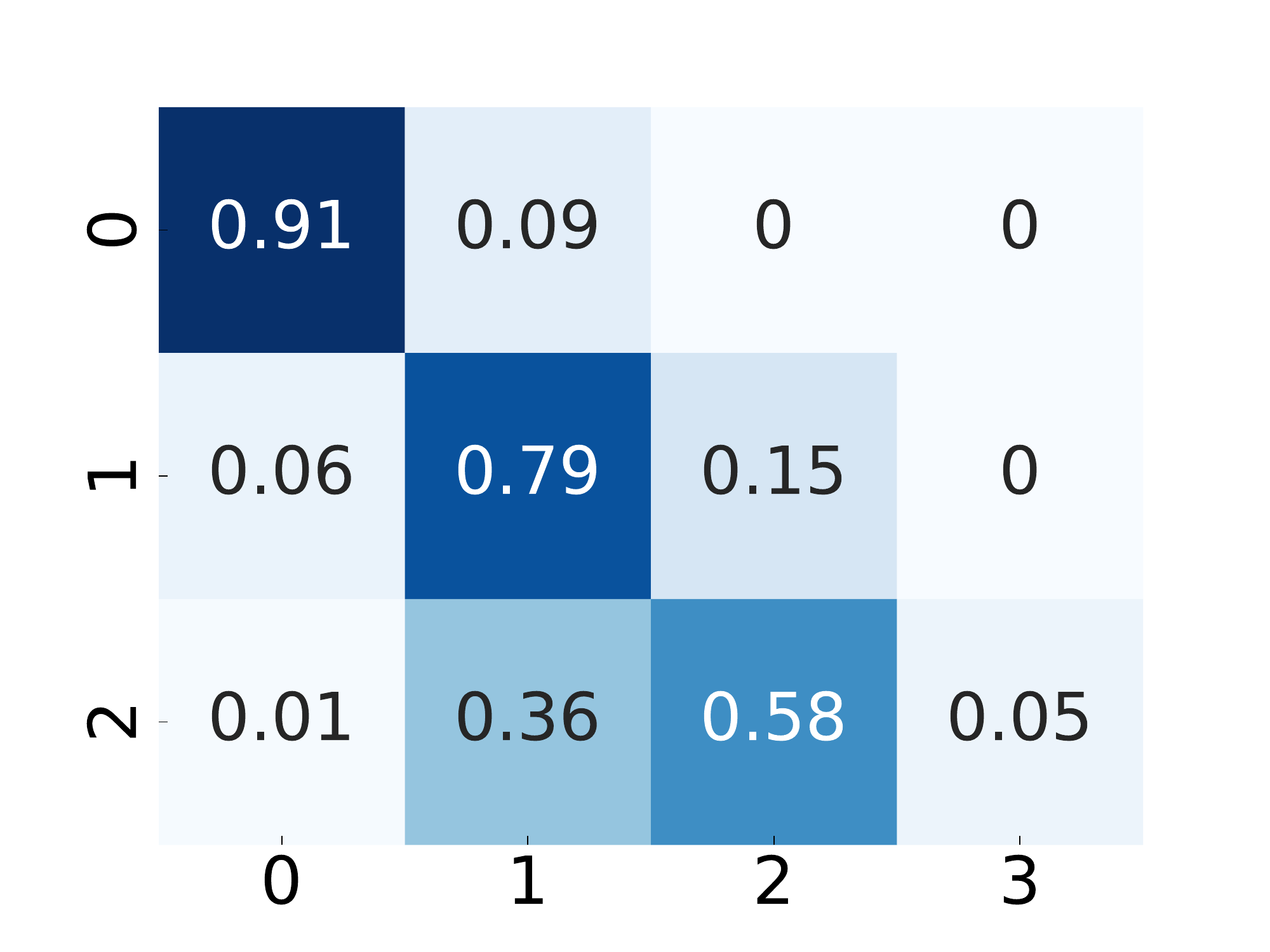}}}%
    \vspace{4pt}
    \subfloat[ANSYN $O3$]{{\includegraphics[width=0.45\linewidth, height=4cm, keepaspectratio, trim=1cm 1cm 0cm 2cm,clip]{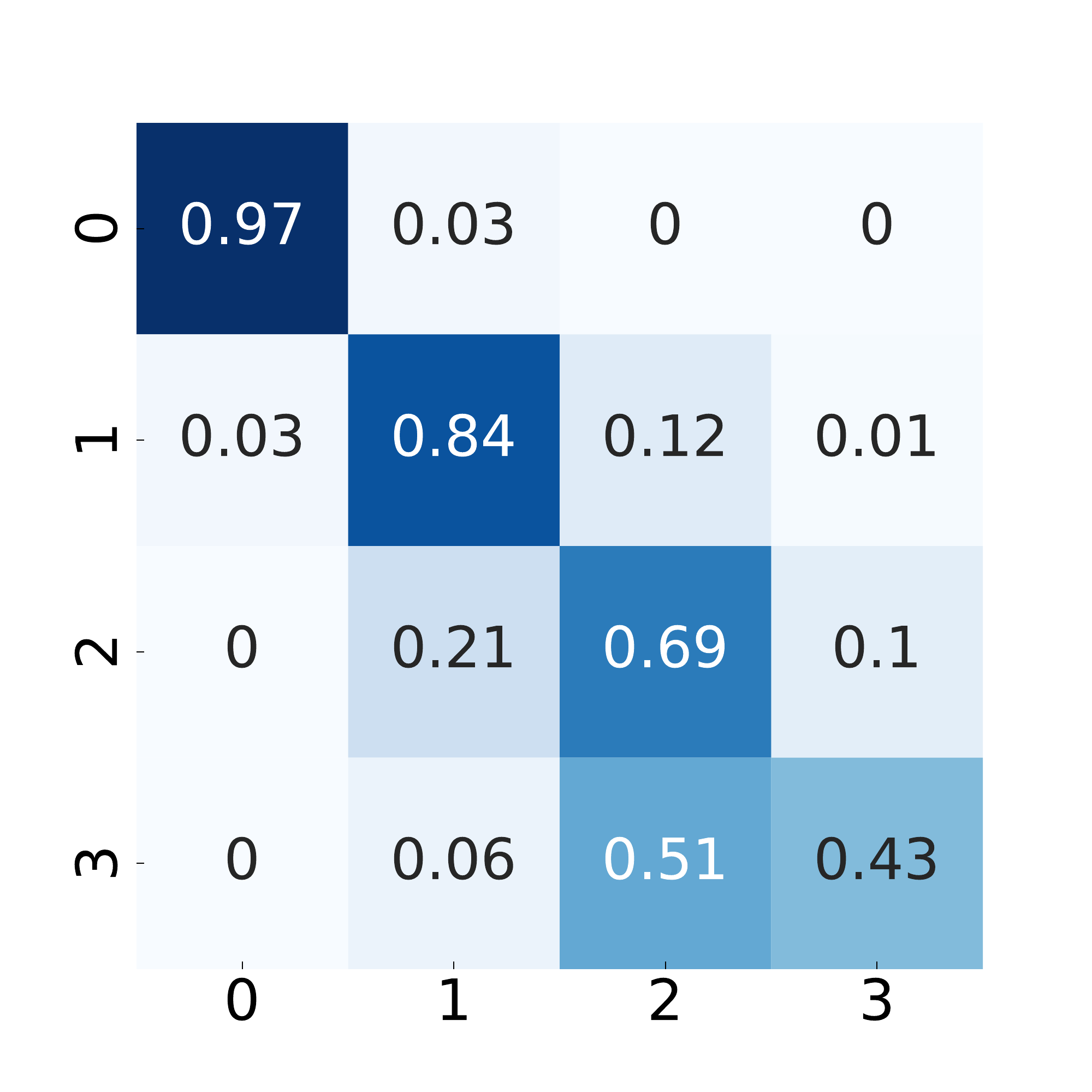}}}%
    \subfloat[RESYN $O3$]{{\includegraphics[width=0.45\linewidth, height=4cm, keepaspectratio, trim=2cm 1.25cm 3cm 2cm,clip]{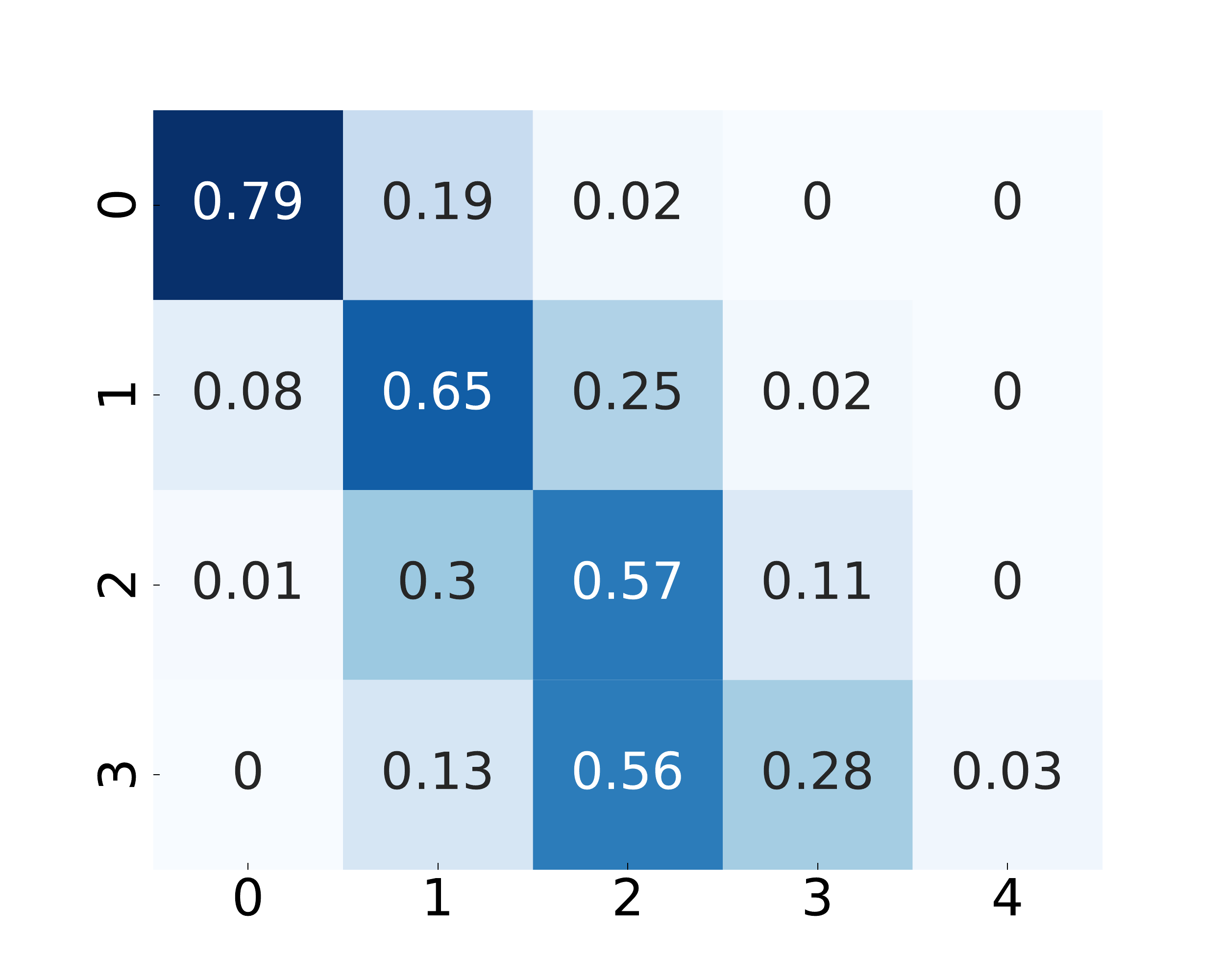}}}%
\caption{Confusion matrix for the number of sound event classes estimated to be active per frame by the SELDnet for ANSYN and RESYN datasets. The horizontal axis represents the SELDnet estimate, and the vertical axis represents the reference.}
\label{fig:conf_mat}

\end{figure}


\subsubsection{Selecting SELDnet output format}
In the output data formats study, it was observed that using the Cartesian $x, y, z$ format in place of azimuth/elevation angle was truly helping the network learn better across datasets as seen in Figure~\ref{fig:xyz_azi_ele}. This suggests that the discontinuity at the angle wrap-around boundary actually reduces the performance of DOA estimation and hence the SELD score.

\subsubsection{Continuous DOA estimation and performance on unseen DOA values} \label{sssec:cont_doa_est}
The input and outputs of SELDnet trained on ANSYN $O1$ and $O2$ subsets for a respective 1000 frame test sequence are visualized in Figure~\ref{fig:crnn_op_visual}. The Figure represents each sound event class and its associated DOA outputs with a unique color. In the case of ANSYN $O1$, we see that the network predictions of SED and DOA are almost perfect. In the case of unseen DOA values (\texttimes{} markers), the network predictions continue to be accurate. This shows that the regression mode output format helps the network learn continuous DOA values, and further that the network is robust to unseen DOA values. In case of ANSYN $O2$, the SED predictions are accurate, while the DOA estimates, in general, are seen to vary around the respective mean reference value. In this work, the DOA and SED labels for a single sound event instance are considered to be constant for the entire duration even though the instance has inherent magnitude variations and silences within. From Figure~\ref{fig:ov2_results} it seems that these variations and silences are leading to fluctuating DOA estimates, while the SED predictions are unaffected. In general, we see that the proposed method successfully recognizes, localizes in time and space, and tracks multiple overlapping sound events simultaneously.

Table~\ref{T:ansim_resim_results} presents the evaluation metric scores for the SELDnet and the baseline methods with ANSYN and RESYN datasets. In the SED metrics for the ANSYN datasets, the SELDnet performed better than the best baseline MSEDnet for $O1$ subset while MSEDnet performed slightly better for $O2$ and $O3$ subsets. With regard to DOA metrics, the SELDnet is significantly better than the baseline DOAnet in terms of frame recall. This improvement in frame recall is a direct result of using SED output as a confidence measure for estimating DOA, thereby extending state-of-the-art SED performance to SELD. Although the frame recall of DOAnet is poor, its DOA error for $O1$ and $O2$ subsets is observed to be lower than SELDnet. The DOA error of the parametric baseline MUSIC with the knowledge of the number of sources is seen to be the best among the evaluated methods for $O2$ and $O3$ subsets. 


\begin{table*}[!htb]
\centering
\caption{SED and DOA estimation metrics for REAL, REALBIG and REALBIGAMB datasets. Best scores for subsets in bold.}
\label{T:real_score}
\begin{tabular}{l|l|ccc|ccc|ccc|ccc|ccc|}
\multicolumn{2}{c|}{} & \multicolumn{3}{c|}{REAL} & \multicolumn{3}{c|}{REALBIG} & \multicolumn{3}{c|}{REALBIGAMB 20dB} & \multicolumn{3}{c|}{REALBIGAMB 10dB} & \multicolumn{3}{c|}{REALBIGAMB 0dB} \\ \cline{3-17}
\multicolumn{1}{c|}{} &Overlap & 1 & 2 & 3 & 1 & 2 & 3 & 1 & 2 & 3 & 1 & 2 & 3 & 1 & 2 & 3 \\ \cline{2-17}
\multicolumn{17}{l}{SED metrics} \\ \hline
SELDnet & ER & 0.40 & 0.49 & 0.53 & 0.37 & 0.42 & 0.50 & \bf0.34 & 0.46 & 0.52 &\bf 0.37 & 0.49 & 0.52 & \bf0.46 & 0.58 & 0.59 \\
 & F & 60.3 & 53.1 & 51.1 & 65.4 & 61.5 & 56.5 & 65.6 & 58.5 & 55.0 & \bf66.3 & 55.4 & 53.3 &\bf 57.9 & 48.6 & 49.0 \\\cline{2-17}
MSEDnet~\cite{Adavanne2017} & ER & \bf 0.35 & \bf0.38 & \bf0.41 &\bf 0.34 &\bf 0.39 &\bf 0.38 & 0.35 & \bf0.40 & \bf0.41 & 0.38 & \bf0.43 & \bf0.42 & 0.48 & \bf0.56 & \bf0.54 \\ 
 & F & \bf66.2 & \bf61.6 & \bf59.5 & 67.3 & 61.8 & 61.9 & \bf66.0 & \bf61.6 & \bf60.1 & 63.2 & \bf58.7 & \bf59.3 & 54.5 & \bf49.3 & \bf51.3 \\\cline{2-17}
SEDnet~\cite{Adavanne2017} & ER & 0.38 & 0.42 & 0.43 & 0.38 & 0.43 & 0.44 & 0.39 & 0.42 & 0.43 & 0.41 & 0.44 & 0.46 & 0.51 & 0.61 & 0.57 \\
 & F & 64.6 & 61.5 & 57.2 & \bf68.0 & \bf62.4 & \bf62.4 & 65.7 & 60.1 & 59.2 & 62.7 & 56.3 & 56.9 & 52.6 & 46.0 & 50.4 \\
\multicolumn{17}{l}{DOA metrics} \\ \hline
SELDnet & DOA error & 26.6 & 33.7 & 36.1 & 23.1 & 31.3 & 34.9 & 25.4 & 32.5 & 36.1 & 27.2 & 32.5 & 36.1 & 30.7 & 33.7 & 36.7 \\
& Frame recall &\bf 64.9 & \bf41.5 &\bf 24.6 & \bf68.0 & \bf45.2 &\bf 28.3 &\bf 69.1 & \bf42.8 & \bf25.8 &\bf 66.9 &\bf 40.0 & \bf27.3 & \bf62.5 & \bf35.2 & \bf23.4 \\ \cline{2-17}
DOAnet~\cite{Adavanne2018_EUSIPCO} & DOA error & \bf6.3 & \bf20.1 & \bf25.8 & \bf7.5 &\bf 17.8 &\bf 22.9 & \bf6.3 & \bf18.9 &\bf 25.78 & \bf8.0 & \bf20.1 & \bf24.1 & \bf14.3 & \bf24.1 & \bf27.5 \\
& Frame recall & 46.5 & 11.5 & 2.9 & 44.1 & 12.5 & 3.1 & 34.7 & 11.6 & 3.2 & 42.1 & 13.5 & 3.3 & 30.1 & 10.5 & 2.8 \\ \cline{2-17}
MUSIC & DOA error & 36.3 & 49.5 & 54.3 & 35.8 & 49.6 & 53.8 & 54.5 & 56.1 & 61.3 & 51.6 & 54.5 & 62.6 & 41.9 & 47.5 & 62.3 
\end{tabular}

\end{table*}

\subsubsection{Performance on mismatched reverberant dataset}
From Table~\ref{T:ansim_resim_results} results on RESYN room 1 subsets, we see that the performance of parametric method MUSIC is poor in comparison to SELDnet in reverberant conditions. The SELDnet is seen to perform significantly better than the baseline DOAnet in terms of frame recall, although the DOAnet has lower DOA error for $O1$ and $O2$ subsets. The SED metrics of SELDnet are comparable if not better than the best baseline performance of MSEDnet. Further, on training the SELDnet on room 1 dataset and testing on moderately mismatched reverberant room 2 and 3 datasets the SED and DOA metric trends remain similar to the results of room 1 testing split. That is, the SELDnet has higher frame recall, the DOAnet has better DOA error, the MUSIC performs poorly, and the SED metrics of SELDnet are comparable to MSEDnet. These results prove that the SELDnet is robust to reverberation in comparison to baseline methods and performs seamlessly on moderately mismatched room configurations.

Figure~\ref{fig:conf_mat} visualizes the confusion matrices for the estimated number of sound event classes per frame by SELDnet. For example in Figure~\ref{fig:ansyn_o2} the SELDnet correctly estimated the number of sources to be two in 76\% (true positive percentage) of the frames which had two sources in the reference. In context, the frame recall value used as a metric to evaluate DOA estimation represents this confusion matrix in one number. From the confusion matrices, we observe that the percentage of true positives drops with higher number of sources, and this drop is even more significant in the reverberant scenario. But, in comparison to the frame recall metric of the baseline DOAnet in Table~\ref{T:ansim_resim_results}, the SELDnet frame recall is significantly better for higher number of overlapping sound events, especially in the reverberant conditions.

\subsubsection{Performance on the size of the dataset}\label{sssec:data_size}
The overall performance of SELDnet on REAL dataset (Table~\ref{T:real_score}) reduced in comparison to ANSYN and RESYN datasets. The baseline MSEDnet is seen to perform better than SELDnet in terms of SED metrics. Similar performance drop on real-life datasets has also been reported on SED datasets in other studies~\cite{emre_TASLP2016}. With regard to DOA metrics, the frame recall of SELDnet continues to be significantly better than DOAnet, while the DOA error of DOAnet is lower than SELDnet. The performance of MUSIC is seen to be poor in comparison to both DOAnet and SELDnet. With the larger REALBIG dataset the SELDnet performance was seen to improve. A similar study was done with larger ANSYN and RESYN datasets, where the results were comparable with that of smaller datasets. This shows that the datasets with real-life IR are more complicated than synthetic IR datasets, and having larger real-life datasets helps the network learn better.


\subsubsection{Performance with ambiance at different SNR}
In presence of ambiance, SELDnet was seen to be robust for 10 and 20 dB SNR REALBIGAMB datasets (Table~\ref{T:real_score}). In comparison to the SED metrics of REALBIG dataset with no ambiance, the SELDnet performance on $O1$ subsets of 10 dB and 20 dB ambiance is comparable, while a small drop in performance was observed with the respective $O2$ and $O3$ subsets. Whereas, the performance was seen to drop considerably for the 0 dB SNR dataset. With respect to DOA error, the SELDnet performed better than MUSIC but poorer than DOAnet across datasets, on the other hand, SELDnet gave significantly higher frame recall than DOAnet. From the insight of SELDnet performance on REAL dataset (Section~\ref{sssec:data_size}), the more complex the acoustic scene the larger the dataset size required to learn better. Considering that the SELDnet is jointly estimating the DOA along with SED in a challenging acoustic scene with ambiance the SELDnet performance can potentially improve with larger datasets.

\begin{table}[!tp]
\centering
\caption{SED and DOA estimation metrics for CANSYN and CRESYN datasets. Best scores for subsets in bold.}
\label{T:circ_array_results}
\resizebox{\columnwidth}{!}{

\begin{tabular}{l|l|ccc|ccc}
\multicolumn{2}{l|}{}  & \multicolumn{3}{c|}{CANSYN}& \multicolumn{3}{c}{CRESYN} \\ \cline{3-8}
 & Overlap & 1 & 2 & 3 & 1 & 2 & 3 \\ \cline{2-8}
\multicolumn{8}{l}{SED metrics} \\ \hline
SELDnet & ER & 0.11 & \bf0.18 & 0.19 & 0.13 & 0.22 & 0.30\\
 & F score & 93.0 & 86.6 & 85.3 & 90.4 & 82.2 & 78.0 \\\cline{2-8}
SELDnet-azi & ER &\bf 0.08 & 0.19 & 0.24  & \bf0.06 & \bf0.18 & \bf0.20\\
 & F score & \bf94.7 & 87.5 & 83.8  & \bf96.3 & \bf87.9 & \bf85.6\\\cline{2-8}
MSEDnet~\cite{Adavanne2017} & ER & 0.09 &\bf 0.18 &\bf 0.16  & 0.12 & 0.22 & 0.26\\
 & F score & 94.6 &\bf 89.0 & \bf86.7  & 92.7 & 83.7 & 80.7\\\cline{2-8}
SEDnet~\cite{Adavanne2017} & ER & 0.15 & 0.21 & 0.20  & 0.18 & 0.26 & 0.25\\
 & F score & 91.4 & 87.3 & 84.7  & 90.5 & 84.3 & 82.8\\\cline{2-8}
HIRnet~\cite{Hirvonen2015} & ER & 0.41 & 0.45 & 0.62  & 0.43 & 0.46 & 0.50\\
 & F score & 60.0 & 54.9 & 58.8 & 59.3 & 60.2 & 58.6 \\
\multicolumn{8}{l}{DOA metrics} \\ \hline
SELDnet & DOA error & 29.5 & 31.3 & 34.3 & 28.4 & 33.7 & 41.0 \\
 & Frame recall & 97.9 & 78.8 & 67.0  & 96.4 & 75.7 & 60.7\\\cline{2-8}
SELDnet-azi & DOA error & 7.5 & 14.4 & 19.6 & 5.2 & 13.2 & 18.4 \\
 & Frame recall & 98.0 &\bf 82.1 & \bf66.2  &\bf 98.5 & \bf82.3 &\bf 70.6\\\cline{2-8}
HIRnet~\cite{Hirvonen2015} & DOA error & 5.2 & 16.3  & 33.0 & 7.4 & 18.6 & 43.3 \\
 & Frame recall & 60.2 & 35.9 & 18.4  & 56.9 & 20.5 & 10.7 \\\cline{2-8}
AZInet~\cite{Chakrabarty2017_nips} & DOA error & \bf1.2 & \bf4.0 & 7.4&\bf 2.3 & \bf6.9 & \bf9.7 \\
 & Frame recall & \bf99.4 & 80.5 & 60.5 & 97.3 & 65.2 & 44.8 
\end{tabular}
}
\end{table}


\subsubsection{Generic to array structure}
The results on circular array datasets are presented in Table~\ref{T:circ_array_results}. With respect to SED metrics, the SELDnet-azi performance is seen to be better than the best baseline MSEDnet for all subsets of CRESYN dataset, while MSEDnet is seen to perform better for $O2$ and $O3$ subsets of CANSYN dataset. Similarly, in the case of DOA metrics, the SELDnet-azi has better frame recall than the best baseline method AZInet across datasets (except for CANSYN $O1$). Whereas, AZInet has lower DOA error than SELDnet-azi. Between SELDnet and SELDnet-azi, even though the frame recall is in the same order the DOA error of SELDnet-azi are lower than SELDnet. This shows that estimating DOA in 3D ($x,y,z$) is challenging using a circular array. Overall, the SELDnet is shown to perform consistently across different array structures (Ambisonic and circular array), with good results in comparison to baselines. 

The usage of SED output as a confidence measure for estimating the number of DOAs in the proposed SELDnet is shown to improve the frame recall significantly and consistently across the evaluated datasets. On the other hand, the DOA error obtained with SELDnet is consistently higher than the classification based baseline DOA estimation methods~\cite{Chakrabarty2017_nips, Adavanne2018_EUSIPCO}. We believe that this might be a result of the regression-based DOA estimation approach in SELDnet not having completely learned the full mapping between input feature and the continuous DOA space. The investigation of which is planned for future work. In general, a classification only or a classification-regression based SELD approach can be chosen based on the required frame recall, DOA error, resolution of DOA labels, training split size, and robustness to unseen DOA values and reverberation.


\section{Conclusion}
\label{sec:conclusion}
In this paper, we proposed a convolutional recurrent neural network (SELDnet) to simultaneously recognize, localize and track sound events with respect to time. The localization is done by estimating the direction of arrival (DOA) on a unit sphere around the microphone using 3D Cartesian coordinates. We tie each sound event output class in the SELDnet to three regressors to estimate the respective Cartesian coordinates. We show that using regression helps estimating DOA in a continuous space, and also estimating unseen DOA values accurately. On the other hand, estimating a single DOA for each sound event class does not allow recognizing multiple instances of the same class overlapping. We plan to tackle this problem in our future work. 

The usage of SED output as a confidence measure to estimate DOA was seen to extend the state-of-the-art SED performance to SELD resulting in a higher recall of DOAs. With respect to the estimated DOA error, although the classification based baseline methods had poor recall they had lower DOA error in comparison to the proposed regression based DOA estimation. The proposed SELDnet uses phase and magnitude spectrogram as the input feature. The usage of such non-method-specific feature makes the method generic and easily extendable to different array structures. We prove this by evaluating on datasets of Ambisonic and circular array format. The proposed SELDnet is shown to be robust to reverberation, low SNR scenarios and unseen rooms with comparable room-sizes. Finally, the overall performance on dataset synthesized using real-life impulse response (IR) was seen to drop in comparison to artificial IR dataset, suggesting the need for larger real-life training datasets and more powerful classifiers in future. 

\bibliographystyle{IEEEtran}
\bibliography{refs}

\begin{IEEEbiography}[{\includegraphics[width=1in,height=1.25in,clip,keepaspectratio]{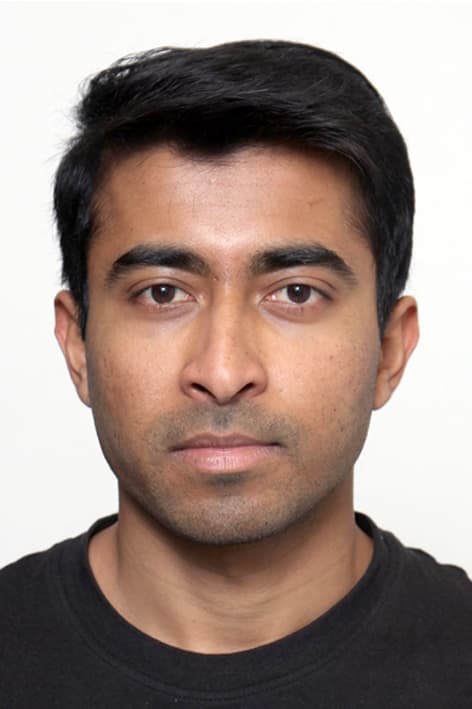}}]{Sharath Adavanne} received his M.Sc. degree in Information Technology from Tampere University of Technology (TUT), Finland in 2011. From 2011 to 2016 he worked in the industry solving problems related to music information retrieval, speech recognition, audio fingerprinting and general audio content analysis. Since 2016, he is pursuing his Ph.D. degree at the laboratory of signal processing in TUT. His current research interest is in the application of machine learning based methods for real-life auditory scene analysis.
\end{IEEEbiography}

\begin{IEEEbiography}[{\includegraphics[width=1in,height=1.25in,clip,keepaspectratio]{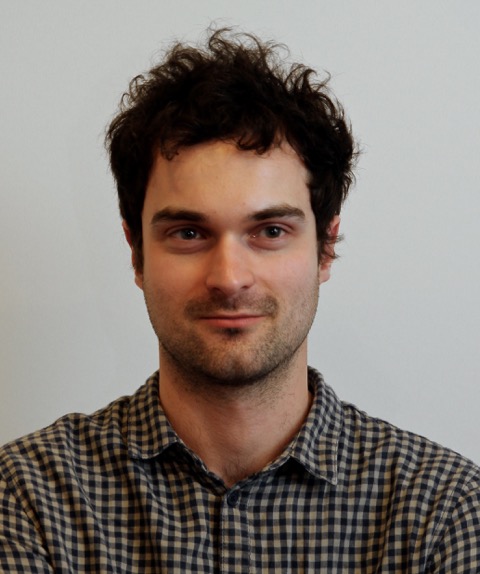}}]{Archontis Politis} obtained a M.Eng degree in civil engineering Aristotle University, Thessaloniki, Greece, and his M.Sc degree in sound and vibration studies from Institute of Sound and Vibration Studies (ISVR), Southampton, UK, in 2006 and 2008 respectively. From 2008 to 2010 he worked as a graduate acoustic consultant in Arup Acoustics, UK, and as a researcher in a joint collaboration between Arup Acoustics and the Glasgow School of Arts, on architectural auralization using spatial sound techniques. In 2016 he obtained a Doctor of Science degree on the topic of parametric spatial sound recording and reproduction from Aalto University, Finland. He has also completed an internship at the Audio and Acoustics Research Group of Microsoft Research, during summer of 2015. He is currently a post-doctoral researcher at Aalto University. His research interests include spatial audio technologies, virtual acoustics, array signal processing and acoustic scene analysis.\end{IEEEbiography}

\begin{IEEEbiography}[{\includegraphics[width=1in,height=1.25in,clip,keepaspectratio]{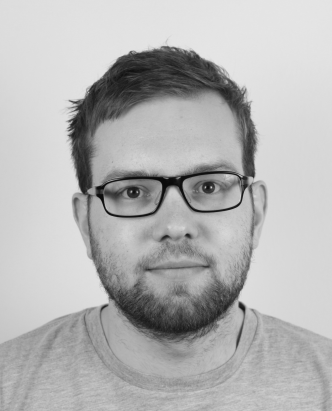}}]{Joonas Nikunen} received the M.Sc degree in signal processing and communications engineering and Ph.D degree in Signal Processing from Tampere University of Technology (TUT), Finland in 2010 and 2015, respectively. He is currently a post-doctoral researcher at TUT focusing on sound source separation with applications on spatial audio analysis, modification and synthesis. His other research interests include microphone array signal processing, 3D/360 audio in general and machine and deep learning for source separation.
\end{IEEEbiography}

\begin{IEEEbiography}[{\includegraphics[width=1in,height=1.25in,clip,keepaspectratio]{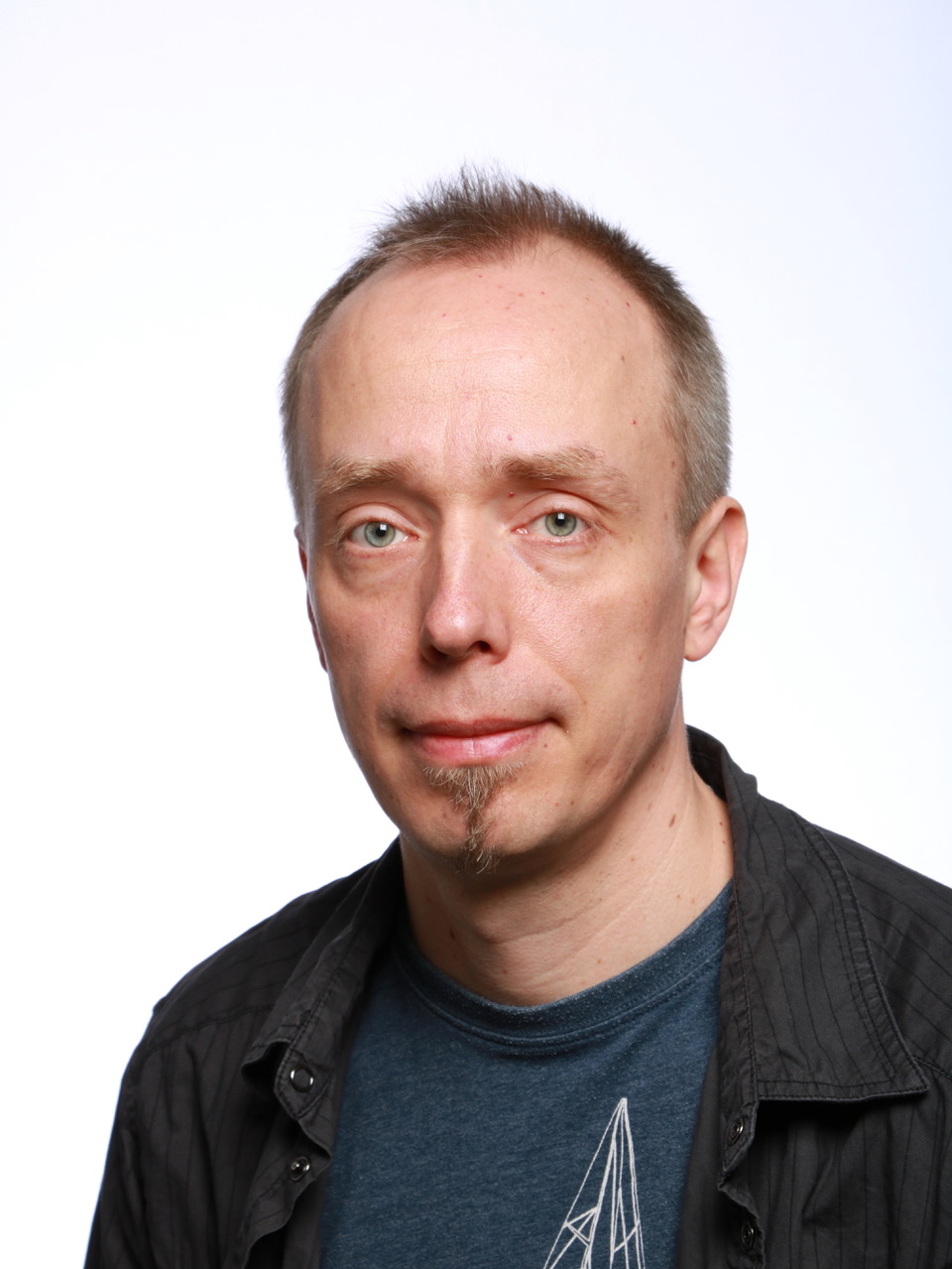}}]{Tuomas Virtanen} is Professor at Laboratory of Signal Processing, Tampere University of Technology (TUT), Finland, where he is leading the Audio Research Group. He received the M.Sc. and Doctor of Science degrees in information technology from TUT in 2001 and 2006, respectively. He has also been working as a research associate at Cambridge University Engineering Department, UK. He is known for his pioneering work on single-channel sound source separation using non-negative matrix factorization based techniques, and their application to noise-robust speech recognition and music content analysis. Recently he has done significant contributions to sound event detection in everyday environments. In addition to the above topics, his research interests include content analysis of audio signals in general and machine learning. He has authored more than 150 scientific publications on the above topics, which have been cited more than 6000 times. He has received the IEEE Signal Processing Society 2012 best paper award for his article "Monaural Sound Source Separation by Nonnegative Matrix Factorization with Temporal Continuity and Sparseness Criteria" as well as three other best paper awards. He is an IEEE Senior Member, member of the Audio and Acoustic Signal Processing Technical Committee of IEEE Signal Processing Society, Associate Editor of IEEE/ACM Transaction on Audio, Speech, and Language Processing, and recipient of the ERC 2014 Starting Grant.
\end{IEEEbiography}

\end{document}